\documentclass[prd,aps,floatfix,preprint,nofootinbib]{revtex4}
\usepackage{amsmath,amssymb,amsbsy,color}
\usepackage{longtable}
\usepackage{epsfig}
\usepackage{graphicx}
\usepackage{amssymb}
\usepackage{epstopdf}
\newcommand{\msbar}{\overline {\rm MS}}
\newcommand{\SF}{Schr\"odinger functional\ }
\newcommand{\no}{\nonumber}
\newcommand{\simg}{\rlap{\raise -4pt \hbox{$\sim$}}
                   \raise 3pt \hbox{$>$}}
\newcommand{\siml}{\rlap{\raise -4pt \hbox{$\sim$}}
                   \raise 3pt \hbox{$<$}}

\begin{document}
 \vspace*{-20mm}
 \begin{flushright}
  \normalsize
  HUPD-1005,\
  KANAZAWA-10-10,\
  KEK-CP-240
 \end{flushright}
\title{
Running coupling constant of ten-flavor QCD with the Schr\"odinger
functional method}

\author{M.~Hayakawa$^{a}$}
\author{K.-I.~Ishikawa$^{b}$}
\author{Y.~Osaki$^{b}$}
\author{S.~Takeda$^{c}$}
\author{S.~Uno$^{a}$}
\author{N.~Yamada$^{d,e}$}
\email[]{norikazu.yamada@kek.jp}

\affiliation{
$^a$ Department of Physics, Nagoya University, Nagoya 464-8602, Japan\\
$^b$ Department of Physics, Hiroshima University, Higashi-Hiroshima
     739-8526, Japan\\
$^c$ School of Mathematics and Physics, College of Science and
     Engineering, Kanazawa university, Kakuma-machi, Kanazawa, Ishikawa 
     920-1192, Japan\\
$^d$ KEK Theory Center, Institute of Particle and Nuclear Studies, High
     Energy Accelerator Research Organization (KEK), Tsukuba 305-0801,
     Japan\\ 
$^e$ School of High Energy Accelerator Science, The Graduate University
     for Advanced Studies (Sokendai), Tsukuba 305-0801, Japan
}

\date{\today}

\begin{abstract}
 
 Walking technicolor theory attempts to realize electroweak symmetry
 breaking as the spontaneous chiral symmetry breakdown caused by the
 gauge dynamics with slowly varying gauge coupling constant and large
 mass anomalous dimension.
 Many-flavor QCD is one of the candidates owning these features.
 We focus on the SU(3) gauge theory with ten flavors of massless
 fermions in the fundamental representation, and compute the gauge
 coupling constant in the \SF scheme.
 Numerical simulation is performed with $O(a)$-unimproved lattice
 action, and the continuum limit is taken in linear in lattice spacing.
 We observe evidence that this theory possesses an infrared fixed
 point.
\end{abstract}

\pacs{12.38.Gc\vspace{-0ex}}
\maketitle

\section{Introduction}
\label{sec:introduction}

While the standard model has been established through a number of
experiments, unnatural hierarchies are present between the electroweak
scale and the Planck scale and also among the fermion masses.
Large Hadron Collider (LHC) is expected to give a new insight into these
hierarchies.
Among various new physics models proposed so far, Technicolor
(TC) model~\cite{Weinberg:1975gm} is one of the most attractive
ones in these regards, as it does not require any fundamental scalar
particles, which cause the former hierarchy, and its extension, Extended
TC model~\cite{Eichten:1979ah}, has a possibility to generate the Yukawa
hierarchy in a dynamical way.
For recent review articles, see, for example,
Refs.~\cite{Chivukula:2000mb}.

TC should be a strongly coupled vector-like gauge system, which triggers
spontaneous chiral symmetry breaking (S$\chi$SB).
It is widely known, however, that the simplest TC models obtained by
rescaling ordinary QCD have already been ruled out by the
$S$-parameter~\cite{Peskin:1990zt} and FCNC~\cite{Chivukula:2010tn}
constraints.
Refs.~\cite{Holdom:1981rm} suggested a series of TC models to circumvent
the FCNC problem.
Those TC models appeal to the gauge dynamics in which the effective
gauge coupling constant runs slowly ({\it i.e.} ``walks'') at a
relatively large value over a wide range of energy scale above the
S$\chi$SB scale, and in which the chiral condensate gets large anomalous
dimension.
Such TC is called {\it walking TC (WTC)}, and possible candidates have
been enumerated through semi-quantitative
analyses~\cite{Dietrich:2006cm}.
Since the dynamics that underlie WTC significantly differ from those of
two or three-flavor QCD, the naive scaling argument in $N_c$ or $N_f$ to
estimate the $S$-parameter would not work, and any quantitative
predictions from WTC require solving nonperturbative dynamics
explicitly.
Lattice gauge theory provides a unique way to study this class of models
from the first principles at present.

Search for candidate theories of WTC is frequently linked to the
$N_f$-dependent phase structure of the gauge theories.
Let us take SU(3) gauge theory with $N_f$ flavors of fermions in the
fundamental representation as an example.
According to the analysis of the perturbative $\beta$-function, the
system with large enough $N_f$ ($N_f > 16.5$) is asymptotically non-free
and trivial unless non-trivial ultraviolet fixed point exists.
On the other hand, if $N_f$ is sufficiently small ($N_f\le 3$) the
dynamics is QCD-like and thus in the chirally broken phase.
It is believed that for the in-between $N_f$ there exists a so-called
conformal phase, where the coupling constant reaches an infrared fixed
point (IRFP) without S$\chi$SB set in, but confinement may take
place~\cite{Caswell:1974gg}.
The range of $N_f$ in which the conformal phase is realized is called
conformal window, and is represented by $N_f^{\rm crit}< N_f < 16.5$.
It is then natural to speculate that the gauge dynamics slightly below
$N_f^{\rm crit}$ exhibit the features required for WTC;
slow running of the gauge coupling constant and S$\chi$SB.
The first goal in the search for WTC is thus to identify
$N_f^{\rm crit}$.

In the past years, many groups have used techniques of lattice
simulations to search for $N_f^{\rm crit}$ and/or WTC through hadron
spectrum, eigenvalue distribution of Dirac operator, the behavior of
running coupling constant, or renormalization group analysis of candidate
theories~\cite{Fleming:2008gy}.
For non-lattice studies, see, for example,
Refs.~\cite{Gies:2005as,Braun:2006jd}.
Among various candidates, many flavor
QCD~\cite{Appelquist:2007hu,Appelquist:2009ty,Deuzeman:2008sc,Deuzeman:2009mh,Fodor:2009wk,Hasenfratz:2009ea,Jin:2009mc,Bilgici:2009nm},
sextet QCD~\cite{Shamir:2008pb,DeGrand:2008kx,DeGrand:2009hu,DeGrand:2010na,Fodor:2009ar,Kogut:2010cz},
and two-color adjoint
QCD~\cite{Hietanen:2009az,Hietanen:2008mr,Bursa:2009we,DelDebbio:2009fd,DelDebbio:2010hu,DelDebbio:2010hx}
have been intensively studied.
In this work, we focus on many flavor QCD with $N_c=3$ and fermions in
the fundamental representation.
In a seminal work~\cite{Appelquist:2007hu}, the running coupling
constants were calculated for eight- and twelve-flavor QCD using the \SF
(SF) scheme on the lattice~\cite{Luscher:1992an}.
They concluded that twelve-flavor QCD has an IRFP at
$g_{\rm SF}^{2} \sim 5$ while eight-flavor QCD does not.
In practice, the study of the running coupling alone is supposed to
be unable to fully exclude the possibility of a large IRFP because it
requires lattice simulations at arbitrarily large coupling.
Even worse, the unphysical, bulk first-order phase transition was found
to occur in strong coupling regime of several gauge
theories~\cite{Aoki:2004iq,Nagai:2009ip,DeGrand:2010na}.
In such simulations, there exists an upper limit on the bare coupling at
which lattice calculation is sensible.
Nevertheless, because of the supports from the spectroscopy
studies~\cite{Deuzeman:2008sc,Fodor:2009wk,Jin:2009mc} the conclusion in
Ref.~\cite{Appelquist:2007hu} that the eight-flavor QCD is QCD-like,
{\it i.e.} $N_f^{\rm crit}>8$, seems to be established nowadays.

After the work of Ref.~\cite{Appelquist:2007hu}, one
group~\cite{Deuzeman:2009mh} has presented an evidence of the
conformality of twelve-flavor QCD.
The opposite conclusion, however, has also been reported by the other
groups~\cite{Fodor:2009wk,Jin:2009mc}.
Therefore $N_f^{\rm crit}<12$ is still under debate.
Clearly the observed contradiction must be clarified before going
further.
While in the spectroscopy study of twelve-flavor QCD many sources of
systematic uncertainties due to finite volume, taste breaking, chiral
extrapolation, lack of continuum limit, {\it etc.}, remain to be
quantified, the calculation of the SF coupling constant of
Ref.~\cite{Appelquist:2007hu} appears, at present, to be less ambiguous.
In such a circumstance, we are tempted to explore the dynamics of
ten-flavor QCD.
In this paper, we investigate, as a first step, the running coupling
constant of ten-flavor QCD on the lattice to see whether it shows
conformal behavior.
We find that the running slows down and observe evidence that
this theory possesses an infrared fixed point.

The paper is organized as follows.
In sec.~\ref{sec:remarks}, we give remarks on how we identify IRFP on
the lattice.
Sec.~\ref{sec:ptanalysis} summarizes the coefficients relevant to the
perturbative calculation of the running coupling constant for later
use.
In sec.~\ref{sec:parameter}, the simulation setup including the
definition of the running coupling constant in the \SF scheme is
presented.
In sec.~\ref{sec:analysis}, we describe analysis method and present
the numerical results.
Sec.~\ref{sec:summary} is devoted to the summary and outlook.

\section{Remarks on searching for IRFP on the lattice}
\label{sec:remarks}

Since there exists a subtlety in proving the existence of IRFP with
lattice gauge theory, in this section we briefly explain what is
actually calculated and then give how to identify the existence of
IRFP.
Here we focus on the concept only.
For further details of the calculational and analysis method that we
take, see the following sections.

In this work, we calculate the renormalized coupling constant in \SF
scheme at two different length scales, $L$ and $s \cdot L$.
In practice, this is realized by repeating the calculation on two
different volumes, $l^4$ and $(s\cdot l)^4$, at a common lattice
bare coupling $g_0^2$, where $l=L/a$.
We denote those couplings by $u$ (or $g^2(g_0^2, l)$) and
$g^2(g_0^2,s\cdot l)$, respectively.
Using those, we define the discrete beta function (DBF) by
$B(u,s,l)=1/g^2(g_0^2,s \cdot l)-1/u$, where the rescaling factor $s$ is
arbitrary but is fixed to 2.
If the DBF is free from lattice discretization errors, the sign of this
quantity may directly tell whether the coupling constant increases or
decreases against the scale change by $s$ at the scale $L$, which is
implicitly set by the value of $u$ that we can choose.
Since discretization errors do exist, however, we need to take the
continuum limit.
The $a\rightarrow 0$ limit is taken for a fixed $L$, {\it i.e.} for a
fixed $u$, by varying lattice spacing $a$.
A series of the DBF thus obtained is then a function of $l$, and the
$l=L/a \rightarrow \infty$ limit is expected to give the continuum
limit.
In summary, the DBF is constructed from a pair of lattice volumes
($l^4$, $(s\cdot l)^4$), and choice of larger $l$ results in the DBF
closer to the continuum limit.

In practice, lattice spacing is varied by changing the lattice bare
coupling $g_0^2$.
If $g^2(g_0^2, l_2)$ turns out to be always larger than
$g^2(g_0^2, l_1)$ with $l_2 > l_1$, $B(u,s,l)<0$ should hold for any $l$
and $s>1$.
In this case, the bare coupling at which $g^2(g_0^2,l_1)$ is equal to a
fixed value $u$ becomes small as lattice size $l_1$ increases or one
approaches the continuum limit. 
Thus the $a\rightarrow 0$ limit is realized in the $g_0^2\rightarrow 0$
limit.
This is the case for asymptotically free theories with no IRFP such as
ordinary QCD, and no subtlety is present.
Even if an IRFP exists in such theories, the situation does not change
as long as the input $u$ is smaller than the IRFP, $g^2_{\rm IRFP}$.
In other words, if the DBF extrapolated to $l\rightarrow \infty$ (or
equivalently $1/l\rightarrow 0$) is negative, the limiting value is
interpreted as the continuum limit and the possibility that an IRFP
exists below $u$ is excluded.

When the DBF extrapolated to $l\rightarrow \infty$ is positive,
interpretation of numerical results becomes ambiguous.
In this case, in the vicinity of $1/l=0$,
$g^2(g_0^2,s \cdot l) < g^2(g_0^2,l)$, {\it i.e.} $B(u,s,l)>0$.
Indeed, it happens below $\beta=4.4$ in Fig.~4 of
Ref.~\cite{Appelquist:2009ty}, for example.
Then, one may expect that the $l\rightarrow \infty$ limit is realized by
$g_0^2\rightarrow \infty$ on first sight.
However, recalling $\phi^4$ theory, this expectation turns out to be too
naive.
In $\phi^4$ theory, the continuum limit exists only in the trivial case
unless the theory possesses a non-trivial UV fixed point.
Since the situation is similar to this case, the most plausible
interpretation is that, when $u > g^2_{\rm IRFP}$, the continuum limit
does not exist unless a nontrivial UV fixed point exists.
Since no nontrivial UV fixed points has been established so far, it is
not suitable to call the extrapolated value the continuum limit when it
is positive.
Nevertheless, we can still infer that $u > g^2_{\rm IRFP}$ because no
other possibility remains.

We investigate the sign of the DBF, starting with the weak coupling regime
$u\sim 1$ where the perturbative calculation is reliable and predicts
a negative value.
We keep monitoring the sign of the DBF with increasing $u$.
The identification of the IRFP is then made by sign-flip of the
DBF extrapolated to $l\rightarrow \infty$.
Notice that, when the extrapolated value is positive, the extrapolation
does not make sense and hence we do not insist that the continuum limit
is determined.

\section{Perturbative analysis}
\label{sec:ptanalysis}

We start with defining the $\beta$ function of an effective gauge
coupling constant in a mass-independent renormalization scheme,
which should have the following expansion in the perturbative regime
\begin{eqnarray}
     \beta(g^2(L))
 &=& L\,\frac{\partial\,g^2(L)}{\partial L}
  = b_1\,g^4(L)+b_2\,g^6(L)+b_3\,g^8(L)+b_4\,g^{10}(L)+\cdots,
  \label{eq:betafunc}
\end{eqnarray}
where $L$ denotes a length scale.
The first two coefficients on the right hand side are
scheme-independent, and given by
\begin{eqnarray}
     b_1
 = \frac{2}{(4\pi)^2}\left[ 11 - \frac{2}{3}N_f\right],
&\ \ \ &
     b_2
 = \frac{2}{(4\pi)^4} \left[\, 102 - \frac{38}{3}N_f\,\right].
\label{eq:b1-b2}
\end{eqnarray}
The remaining coefficients are scheme-dependent and known only in the
limited schemes and orders.
The third coefficient takes the following form in the \SF scheme;
\begin{eqnarray}
     b_3^{\rm SF}
 &=& b_3^{\msbar} + \frac{b_2\,c_2^{\theta}        }{2\pi}
                  -
                  \frac{b_1\,(c_3^{\theta}-{c_2^{\theta}}^2)}{8\pi^2},
\label{eq:b3SF}
\end{eqnarray}
where $b_3^{\msbar}$ is a coefficient in the $\msbar$ scheme,
\begin{eqnarray}
     b_3^{\msbar}
 &=& \frac{2}{(4\pi)^6}
     \left[\, \frac{2857}{ 2}
            - \frac{5033}{18}N_f
            + \frac{ 325}{54}N_f^2\,
     \right],
\label{eq:beta-3MS}
\end{eqnarray}
and the calculable quantities $c_2^{\theta}$ and $c_3^{\theta}$ depend
on the spatial boundary condition imposed on the fermion fields in the
SF setup, {\it i.e} so-called $\theta$.
Those for $\theta=\pi/5$ and $c_{2}^{\theta}$ for $\theta=0$ are known
to be~\cite{Bode:1999sm}
\begin{eqnarray}
     c_2^{\theta=\pi/5}
 &=& 1.25563 + 0.039863 \times N_f,\\
     c_3^{\theta=\pi/5}
 &=& ({c_2^{\theta=\pi/5}})^2 + 1.197(10) + 0.140(6)\times N_f
   - 0.0330(2)\times N_f^2,\\ 
     c_2^{\theta=0}
 &=& 1.25563 + 0.022504\times N_f,
\end{eqnarray}
but $c_3^{\theta=0}$ has not been calculated yet.
Although $\theta=0$ is chosen in our simulation as described in
sec.~\ref{sec:parameter}, the coefficients for $\theta=\pi/5$ are used
only to see the situation of conformal windows inferred just from the
perturbative analysis, and the potential size of difference between the
two- and three-loop calculations.

The perturbative estimates of the infrared fixed point (IRFP) for
SU(3) gauge theory with $N_f$ flavors of fundamental fermion are
summarized in Tab.~\ref{tab:p-IRFP}.
We note that in the three-loop perturbative analysis the existence of
IRFP is determined only by the sign of $b_3$, which is
always negative for the range of $N_f$ shown in Tab.~\ref{tab:p-IRFP}.
Therefore, the existence of IRFP as well as its value may be unstable
against including higher orders.
Nevertheless, for $N_f\ge14$ the difference between the two- and
three-loop results is reasonably small, and one may expect that higher
order corrections do not spoil the existence of IRFP or even do not
change its value by much for such a large $N_f$.

According to the analysis based on Schwinger-Dyson equation, S$\chi$SB
is expected to occur when the running coupling constant reaches
$g^{2}\sim \pi^2$ in SU(3) gauge theories~\cite{Appelquist:1988yc}.
In spite of the scheme-dependence of the running coupling constant and
the value of IRFP, those results motivate us to speculate that
ten-flavor QCD may exhibit strongly coupled walking dynamics, and thus
deserves full nonperturbative calculation.
\begin{table}[tb]
\centering
\begin{tabular}{c|ccccccc}
 $N_f$ & 4 & 6 & 8 & 10 & 12 & 14 & 16\\
\hline
 two-loop universal
    &  -    &  -    &  -    & 27.74 & 9.47 & 3.49 & 0.52\\
 three-loop SF with $\theta=\pi/5$
    & 43.36 & 23.75 & 15.52 & 9.45 & 5.18 & 2.43 & 0.47\\
\end{tabular}\\[2ex]
\caption{The perturbative IRFP obtained from the two-loop universal and
 the three-loop SF scheme analyses.}
\label{tab:p-IRFP}
\end{table}

\section{Simulation details}
\label{sec:parameter}

\subsection{\SF}

We employ the \SF (SF) method~\cite{Luscher:1992an} to study the scale
dependence of the running coupling constant.
Unimproved Wilson fermion action and the standard plaquette gauge action
are used without any boundary counter terms as described below.

The SF on the lattice is defined on a four dimensional hypercubic
lattice with a volume $(L/a)^3 \times (T/a)$ in the cylindrical
geometry.
Throughout this work, the temporal extent $T/a$ is chosen to be equal to
the spatial one $L/a$.
Periodic boundary condition in the spatial directions with vanishing phase
factor ($\theta=0$) and Dirichlet one in the temporal direction are
imposed for both gauge ($U_\mu(x)$) and fermion ($\psi(x)$ and
$\bar\psi(x)$) fields.
The boundary values for gauge and fermion fields are represented by
three-by-three color matrices, $C$ and $C'$, and spinors, $\rho$,
$\rho'$, $\bar\rho$ and $\bar\rho'$, respectively.
The partition function of this system is given by
\begin{eqnarray}
 Z_{\rm SF}(C',\bar \rho',\rho'\,; C,\bar \rho,\rho)
=e^{-\Gamma(C',\bar \rho',\rho'\,; C,\bar \rho,\rho)}
=\int D[U,\psi,\bar\psi]
 e^{-S[U,\psi,\bar\psi,C,C',\rho,\rho',\bar \rho,\bar \rho']},
\end{eqnarray}
where $\Gamma$ is the effective action, and
\begin{equation}
S[U,\psi,\bar\psi,C,C',\rho,\rho',\bar \rho,\bar\rho']
=S_g[U,C,C']+S_q[U,\psi,\bar\psi,\rho,\rho',\bar\rho,\bar\rho']. 
\end{equation}
For the pure gauge part, we employ the plaquette action,
\begin{eqnarray}
S_g[U,C,C'] = \frac{\beta}{6}
         \sum_{x}\sum_{\mu=0}^3\sum_{\nu=0}^3
         \bar\delta_{\mu,\nu}w_{\mu,\nu}(x_0)~
         {\rm Tr}\left[1-P_{\mu,\nu}(x) \right],
\label{eq:gauge action}
\end{eqnarray}
where $\beta=6/g_0^2$ denotes the inverse of the bare coupling constant,
$\bar\delta_{\mu,\nu}$=0 when $\mu=\nu$ otherwise 1, and
$P_{\mu,\nu}(x)$ denotes a 1$\times$1 Wilson loop on the $\mu$-$\nu$
plane starting and ending at $x$.
The spatial link variables on the boundaries, the hypersurfaces at
$x_0=0$ and $L/a$, are all set to the diagonal, constant $SU(3)$
matrices as
\begin{eqnarray}
  U_k(x)|_{x_0=0}
= \exp \left[ C \right],&&
  C
= \frac{ia}{L} \left(
  \begin{array}{ccc}
   \eta-\frac{\pi}{3} & 0 & 0 \\
   0 & -\frac{1}{2}\,\eta & 0 \\
   0 & 0 & -\frac{1}{2}\,\eta+\frac{\pi}{3}
  \end{array} \right),
\label{formul:BGF:C0}\\
  \left. U_k(x)\right|_{x_0=L/a}
= \exp \left[ C' \right],&&
  C'
= \frac{ia}{L} \left(
  \begin{array}{ccc}
   -\eta-\pi & 0 & 0 \\
    0 & \frac{1}{2}\,\eta+\frac{\pi}{3} & 0 \\
    0 & 0 & \frac{1}{2}\,\eta+\frac{2\pi}{3}
  \end{array} \right)
\label{formul:BGF:CT},
\end{eqnarray}
where $k=1$, 2, 3, and $\eta$ is parameterizing the gauge boundary
fields.
The weight $w_{\mu,\nu}(x_0)$ in eq.~(\ref{eq:gauge action}) is given by
\begin{eqnarray}
w_{\mu,\nu}(x_0) &=& \left\{\begin{array}{cl}
  c_t            & \mbox{for ($t=0$ or $t=(L/a)-1$) and
                         ($\mu$ or $\nu$=0)}\\ 
  0              & \mbox{for ($t=(L/a)$) and ($\mu$ or $\nu$=0)}\\
  \frac{1}{2}c_s & \mbox{for ($t=0$ or $t=(L/a)$) and
                         ($\mu\ne$0 and $\nu\ne$0)}\\
  1              & \mbox{for all the other cases}
                  \end{array}\right..
\label{eq:wmunu}
\end{eqnarray}
By tuning $c_t$, $O(a)$ errors induced from the boundaries in the time
direction can be removed perturbatively, but in this work we simply take
its tree level values, $c_t=1$.
With this setup, the value of $c_s$ can be arbitrarily chosen because
the spatial plaquettes on the boundaries do not contribute to the
action.
We thus set $c_s=0$.

The fermion fields are described by the unimproved Wilson fermion action,
\begin{eqnarray}
     S_q[U,\psi,\bar\psi]
 &=& N_f \sum_{x,y} \bar{\psi}(x) D(x,y;U) \psi(y)\,
  =  N_f \sum_{x,y} \bar{\psi}^{\rm lat}(x) D^{\rm lat}(x,y;U) \psi^{\rm lat}(y),
   \label{eqn:setup:Sq1}\\
     D^{\rm lat}(x,y;U)
 &=& \delta_{xy} 
    -\kappa \sum_{\mu} 
     \left\{ \left( 1 - \gamma_{\mu} \right)
             U_\mu(x) \delta_{x+\hat{\mu},y}
           + \left( 1 + \gamma_{\mu} \right)
             U^{\dagger} _\mu(x-\hat \mu)
             \delta_{x-\hat{\mu},y}
     \right\},
   \label{eqn:setup:Sq2}
\end{eqnarray}
where
\begin{eqnarray}
    \psi^{\rm lat}(x)
&=& \frac{1}{\sqrt{2\kappa}}\, \psi(x),\ \ \
    \bar \psi^{\rm lat}(x)
 =  \frac{1}{\sqrt{2\kappa}}\, \bar \psi(x),\ \ \
    D^{\rm lat}(x,y;U)
 =  2\kappa\, D(x,y;U)\,.
\end{eqnarray}
The hopping parameter $\kappa$ is related to the bare mass $m_0$ through
$2\,\kappa=1/(am_0 + 4)$.
The dynamical degrees of freedom of the fermion field $\psi(x)$ and
anti-fermion fields $\bar\psi(x)$ reside on the lattice sites $x$ with
$0<x_0<T$.
On both boundaries ($x_0=0$ and $T$), the half of the Dirac components
are set to zero and the remaining components are fixed to some
prescribed values, $\rho$, $\bar\rho$, $\rho'$ and $\bar\rho'$, as
\begin{eqnarray}
 \left.P_+\psi(x)\right|_{x_0=0}=\rho({\bf x}),\ \ \
 \left.P_-\psi(x)\right|_{x_0=0}=0,\\
 \left.P_-\psi(x)\right|_{x_0=T}=\rho'({\bf x}),\ \ \
 \left.P_+\psi(x)\right|_{x_0=T}=0,\\
 \left.\bar\psi(x)P_-\right|_{x_0=0}=\bar\rho({\bf x}),\ \ \
 \left.\bar\psi(x)P_+\right|_{x_0=0}=0,\\
 \left.\bar\psi(x)P_+\right|_{x_0=T}=\bar\rho'({\bf x}),\ \ \
 \left.\bar\psi(x)P_-\right|_{x_0=T}=0,
\end{eqnarray}
where $P_\pm=(1\pm\gamma_0)/2$.
In this work, the boundary values for the fermion fields are set to
zero, {\it i.e.}
\begin{eqnarray}
\rho=\rho'=\bar\rho=\bar\rho'=0.
\end{eqnarray}

\subsection{Definition of the running coupling}

With the gauge boundary conditions (\ref{formul:BGF:C0}) and
(\ref{formul:BGF:CT}), the absolute minimum of the action is given by a
color-electric background field denoted by $B(x)$.
Then, the effective action can be defined as a function of $B$ by
\begin{eqnarray}
 \Gamma[B]=-\ln Z_{\rm SF}(C',\bar \rho',\rho'\,; C,\bar \rho,\rho),
\end{eqnarray}
which has the following perturbative expansion in the bare coupling
constant,
\begin{eqnarray}
 \Gamma=\frac{1}{g_0^2}\,\Gamma_0 + \Gamma_1 + O(g_0^2)\,,
\end{eqnarray}
and, in particular, the lowest-order term
\begin{eqnarray}
 \Gamma_0 = \left[g_0^2\,S_g[B]\right]_{g_0=0},
\end{eqnarray}
is exactly the classical action of the induced background field.
The SF scheme coupling is then defined in the massless limit for
fermions by
\begin{eqnarray}
   \left.\frac{\partial \Gamma}{\partial \eta}\right|_{\eta=0}
 = \frac{1}{g_{\rm SF}^2(g_0^2,\,l=L/a)}\,
   \left.\frac{\partial \Gamma_0}{\partial \eta}\right|_{\eta=0}
 = \frac{k}{g_{\rm SF}^2(g_0^2,\,l)},
\end{eqnarray}
where the normalization constant $k$ is determined such that
$g_{\rm SF}^2=g_0^2$ holds in the leading order of the perturbative
expansion, and is found to be
\begin{eqnarray}
   k
 = \left.\frac{\partial \Gamma_0}{\partial \eta}\right|_{\eta=0}
 = 12 \left(\frac{L}{a}\right)^2
   \left[\sin\left(2\gamma\right)+\sin\left(\gamma\right)
   \right]
 = k
 \ \ \ \mbox{ with }
 \gamma=\frac{\pi}{3}\left(\frac{a}{L}\right)^2.
\end{eqnarray}
Because of the absence of the clover term, only the $\eta$-derivative of
the gauge action contributes to $1 /g_{\rm SF}^2(g_0^2,\,l)$.

\subsection{Parameters}

The simulation was performed on the lattice sizes of
$l^4=(L/a)^{4}$ = $4^4$, $6^{4}$, $8^{4}$, $12^{4}$, and $16^{4}$ with
the inverse of bare gauge coupling constant $\beta=6/g^{2}_{0}$ in the
range, $4.4 \le \beta \le 96.0$.
However, the data from $l=4$ lattices are not used in the following
analysis because it was found that they have large discretization
errors.
We calculated the SF coupling on $18^4$ lattice with a single $\beta$
($\beta$=4.55), and the result is used to check the scaling violation at
a specific value of $g_{\rm SF}^2$.

The algorithm to generate the gauge configuration follows the standard
HMC with five pseudo-fermion fields introduced to simulate the ten
flavors of dynamical fermions.
The numerical simulations were carried out on several different
architectures including GPGPU, PC cluster and supercomputers.
In order to achieve high performance on each architecture, the HMC
code, especially the fermion solver part, were optimized depending on
each architecture.
In particular, mixed precision solver using multiple GPUs enables us
to obtain high statistics on $g_{\rm SF}^2$ at $l^4 = 12^4$ and
$16^4$~\cite{Ishikawa:2010aa}.
Acceptance ratio is kept to around 80 \% by adjusting the molecular
dynamics step size ($\delta \tau$).

Since the Wilson fermion explicitly breaks chiral symmetry, the value of
$\kappa$ is tuned, for every pair of $(\beta,\ L/a)$, to its critical
value $\kappa_c$ realizing the massless fermion by monitoring the
corresponding PCAC mass.
The values of $\beta$, $\kappa$, the number of trajectories,
$\delta \tau$ and the results for $l=L/a$=6, 8, 12, 16, and 18 lattices
are tabulated in
Tabs.~\ref{tab:simpara_L6_imp0}-\ref{tab:simpara_L18_imp0},
respectively.

\subsection{Comment on $O(a)$-unimprovement}

In our pilot study, we employed the $O(a)$-improved fermion action with
the perturbatively determined counter terms.
With this setup, we encountered a sudden change of the plaquette and the
PCAC mass at $l$=6 and $\beta$=3.6 when $\kappa$ was decreased from
0.1517, and we could not realize the vanishing PCAC mass.
The expected SF coupling constant is about $3 \sim 4$ there.
The same phenomenon also occurs on $l=4$ lattices at almost the same
value of bare coupling constant.
Since the observed behavior looks similar to those reported
in Refs.~\cite{Aoki:2004iq,Nagai:2009ip,DeGrand:2010na}, we infer that
this is a bulk, first order phase transition.
In order to cover the region $g_{\rm SF}^2 \sim O(10)$, we omitted any
$O(a)$ improvements.
Thus the leading discretization error in our result is linear in lattice
spacing.

Even without $O(a)$ improvements, the bulk, first order phase transition
is observed for $\beta=6/g_0^2\sim 4.4$.
However, this time it happens at the renormalized coupling constant greater
than the $O(a)$-improved case, typically $g_{\rm SF}^2\sim O(10)$.
Since this bulk phase transition is considered as a lattice artifact,
whenever this happens we discard the gauge configurations at such
$\beta$.
Thus the position of the critical $\beta$ ($\sim 4.4$) sets the lower
limit on our exploration of $\beta$.

\section{Analysis method and Results}
\label{sec:analysis}

\subsection{Raw data}
\label{sec:results}

\begin{figure}[h]
 \centering
 \begin{tabular}{cc}
  \includegraphics*[width=0.5 \textwidth,clip=true]
  {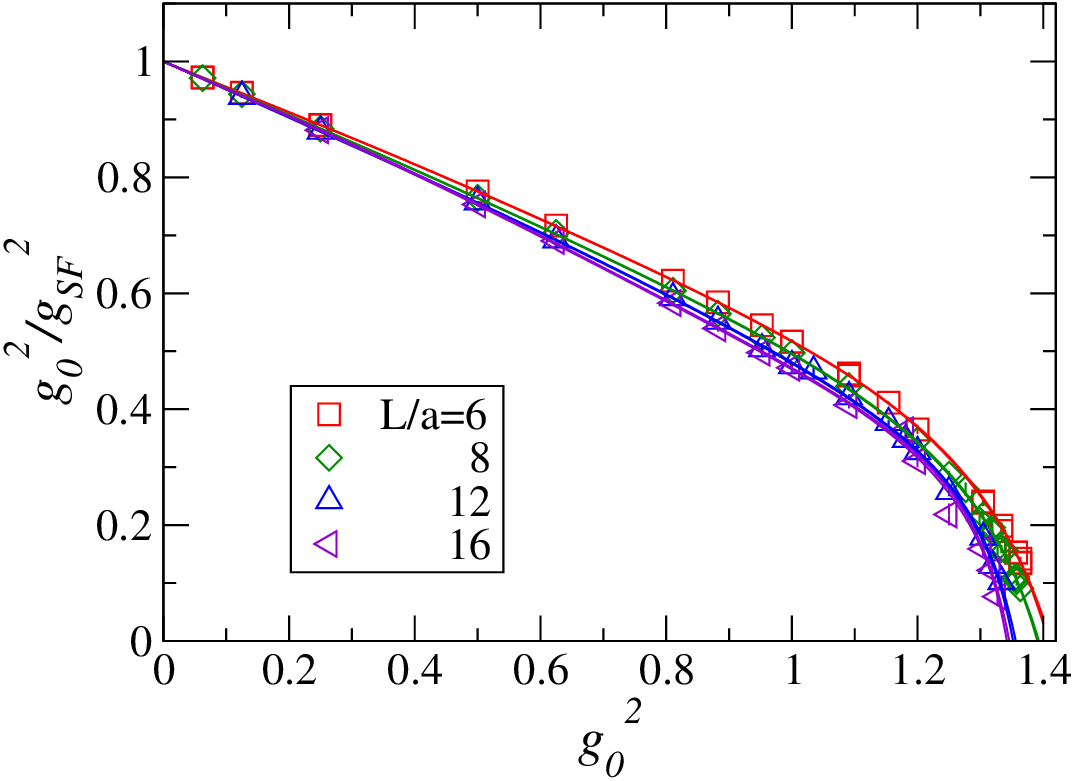}&
  \includegraphics*[width=0.5 \textwidth,clip=true]
  {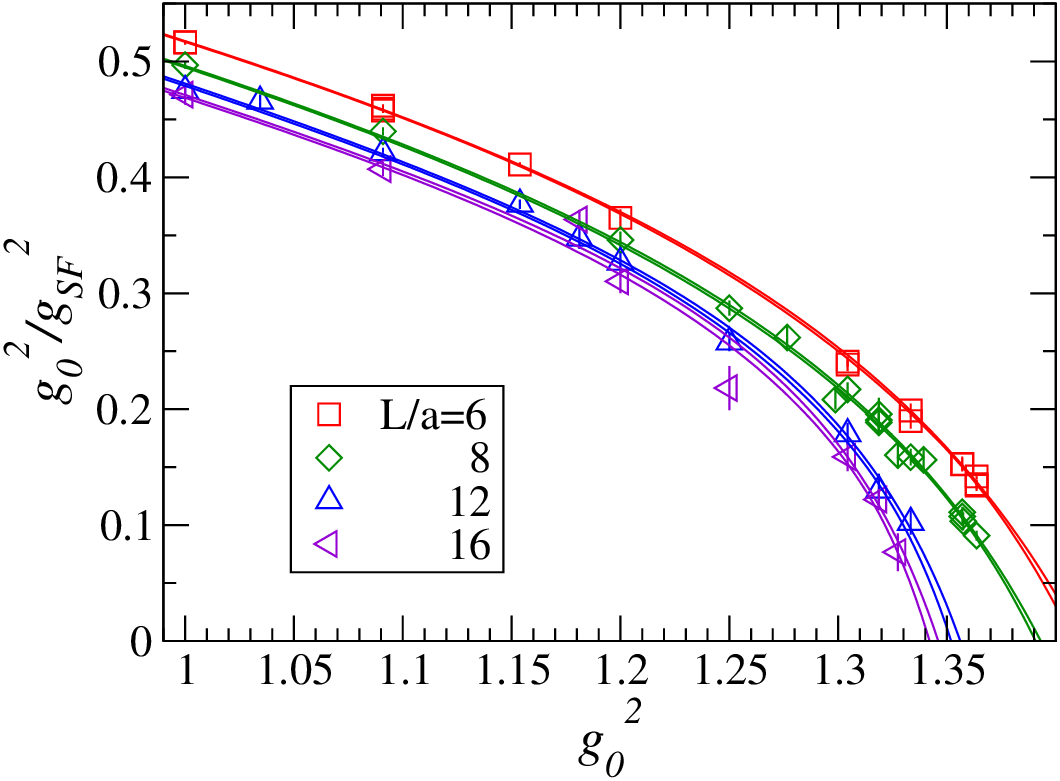}\\
 \end{tabular}
\caption{$g_0^2$ dependence of $g_0^2/g_{\rm SF}^2$ for $l=L/a$=6, 8, 12
 and 16.
 The right panel magnifies the region of $g_0^2 \in [1.1,\ 1.40]$.}
\label{fig:beta_vs_g2}
\end{figure}

The SF coupling constant ($g_{\rm SF}^2$) and the PCAC mass ($M$)
obtained on each ($\beta$, $\kappa$, $l$) are shown in
Tabs.~\ref{tab:simpara_L6_imp0}-\ref{tab:simpara_L18_imp0}.
$g_0^2/g_{\rm SF}^2$ is plotted as a function of the bare coupling
constant $g_0^2$ in Fig.~\ref{fig:beta_vs_g2}.
The figure shows that $g_{\rm SF}^2$ increases with $l=L/a$ at a
fixed $g_0^2$, but the change between the data from $l=12$ and $l=16$ is
tiny.
For later use, we fit the data of $g_0^2/g_{\rm SF}^2$ to an
interpolating formula as a function of the bare coupling constant
$g_0^2$.
Among various functional forms we examined, the following form
\begin{eqnarray}
     \frac{g_0^2}{g_{\rm SF}^2(g^{2}_{0}, l)}
 &=& \frac{1-a_{l,1}\,g_0^4}
          {1+p_{1,l}\times g_0^2+
           \sum_{n=2}^N a_{l,n} \times g_0^{2\,n}
          },
 \label{eq:fitfunc}
\end{eqnarray}
turned out to give the minimum $\chi^2/$dof for a fixed number of free
parameters, $N$.
We thus employ eq.~(\ref{eq:fitfunc}).
In eq.~(\ref{eq:fitfunc}), $p_{1,l}$ is the $l$-dependent coefficient
and we have calculated them perturbatively in the SF scheme
\begin{eqnarray}
 p_{1,l} = \left\{
 \begin{array}{ll}
0.4477107831 & \mbox{ for } l=6 \\
0.4624813408 & \mbox{ for } l=8 \\
0.4756888260 & \mbox{ for } l=12 \\
0.4833079203 & \mbox{ for } l=16 \\
0.4864767958 & \mbox{ for } l=18
 \end{array}
 \right..
\end{eqnarray}
The other coefficients $a_{l,n}$'s are determined for each $l$
independently.
We optimize the degree of polynomial $N$ in the denominator of
eq.~(\ref{eq:fitfunc}) by monitoring $\chi^{2}$/dof, and take $N=5$ for
$l$ = 6 and 12, and $N=4$ for $l$ = 8 and 16.
Tab.~\ref{tab:fit_param1_pade_4} shows the fit results for the
coefficients in eq.~(\ref{eq:fitfunc}).
The fit results are also shown as the region sandwiched by a pair of
solid curves in Fig.~\ref{fig:beta_vs_g2}.
\begin{table}[h]
\begin{tabular}{rccrccccr}
$\beta$ & $\kappa$ & Trajs. & plq. & $\delta\tau$ & Acc. & $g_{\rm SF}^2$ & $M$ \\
\hline
96.0000 & 0.1267030 & 39,700 & 0.979268(0.000002) &0.0076 &0.827(0.002) & 0.06431(0.00006) &$ 0.00012$(0.00003)\\
96.0000 & 0.1267070 & 49,900 & 0.979267(0.000002) &0.0076 &0.826(0.002) & 0.06428(0.00005) &$-0.00004$(0.00003)\\
\hline
48.0000 & 0.1276060 & 40,100 & 0.958852(0.000005) &0.0098 &0.857(0.002) & 0.13221(0.00010) &$-0.00013$(0.00002)\\
48.0000 & 0.1276100 & 41,100 & 0.958846(0.000004) &0.0098 &0.857(0.002) & 0.13209(0.00010) &$-0.00016$(0.00002)\\
\hline
24.0000 & 0.1295180 & 24,700 & 0.917566(0.000009) &0.0149 &0.848(0.002) & 0.28079(0.00015) &$ 0.00006$(0.00002)\\
24.0000 & 0.1295200 & 60,300 & 0.917562(0.000005) &0.0152 &0.838(0.002) & 0.28086(0.00010) &$ 0.00006$(0.00001)\\
\hline
12.0000 & 0.1339640 & 48,700 & 0.833056(0.000014) &0.0250 &0.812(0.002) & 0.64450(0.00076) &$-0.00004$(0.00005)\\
\hline
9.6000 & 0.1365680 & 160,300 & 0.789765(0.000012) &0.0256 &0.826(0.001) & 0.87189(0.00068) &$ 0.00002$(0.00003)\\
\hline
7.4000 & 0.1410690 & 120,500 & 0.724148(0.000015) &0.0270 &0.854(0.001) & 1.30413(0.00194) &$ 0.00006$(0.00006)\\
\hline
6.8000 & 0.1430520 & 120,300 & 0.698517(0.000018) &0.0270 &0.870(0.001) & 1.51024(0.00244) &$-0.00025$(0.00008)\\
\hline
6.3000 & 0.1451400 & 17,400 & 0.673231(0.000076) &0.0333 &0.817(0.001) & 1.74788(0.00505) &$ 0.00015$(0.00019)\\
\hline
6.0000 & 0.1466380 & 33,600 & 0.655993(0.000034) &0.0333 &0.837(0.002) & 1.93684(0.00691) &$ 0.00044$(0.00021)\\
6.0000 & 0.1466410 & 80,300 & 0.655981(0.000029) &0.0333 &0.833(0.001) & 1.93605(0.00362) &$ 0.00004$(0.00010)\\
\hline
5.5000 & 0.1497590 & 50,300 & 0.622923(0.000025) &0.0370 &0.817(0.002) & 2.38340(0.01092) &$ 0.00042$(0.00020)\\
5.5000 & 0.1497610 & 36,000 & 0.622942(0.000027) &0.0357 &0.827(0.002) & 2.36232(0.00634) &$-0.00018$(0.00022)\\
5.5000 & 0.1497620 & 140,300 & 0.622977(0.000023) &0.0357 &0.831(0.001) & 2.37542(0.00963) &$ 0.00023$(0.00014)\\
\hline
5.2000 & 0.1521330 & 220,300 & 0.600097(0.000019) &0.0380 &0.812(0.001) & 2.80668(0.01246) &$-0.00015$(0.00014)\\
\hline
5.0000 & 0.1539800 & 59,900 & 0.583463(0.000049) &0.0400 &0.806(0.002) & 3.28837(0.06618) &$-0.00005$(0.00046)\\
\hline
4.6000 & 0.1585140 & 33,800 & 0.545776(0.000055) &0.0400 &0.813(0.002) & 5.47008(0.13064) &$ 0.00092$(0.00043)\\
4.6000 & 0.1585150 & 150,000 & 0.545680(0.000041) &0.0400 &0.813(0.001) & 5.41263(0.09891) &$ 0.00123$(0.00042)\\
\hline
4.5000 & 0.1599020 & 100,300 & 0.535280(0.000061) &0.0400 &0.813(0.001) & 7.02516(0.24479) &$ 0.00111$(0.00069)\\
4.5000 & 0.1599030 & 100,300 & 0.535305(0.000066) &0.0400 &0.813(0.002) & 6.70575(0.19622) &$ 0.00033$(0.00061)\\
\hline
4.4215 & 0.1610680 & 105,900 & 0.526537(0.000087) &0.0385 &0.825(0.001) & 8.88882(0.36944) &$ 0.00238$(0.00097)\\
4.4215 & 0.1610820 & 92,400 & 0.526692(0.000066) &0.0385 &0.826(0.001) & 8.90139(0.32355) &$ 0.00073$(0.00075)\\
\hline
4.4000 & 0.1614210 & 249,500 & 0.524331(0.000060) &0.0400 &0.811(0.001) & 9.60163(0.19661) &$ 0.00051$(0.00050)\\
4.4000 & 0.1614220 & 182,500 & 0.524342(0.000091) &0.0400 &0.812(0.001) &10.17980(0.33990) &$ 0.00119$(0.00073)\\
4.4000 & 0.1614230 & 250,500 & 0.524387(0.000062) &0.0400 &0.811(0.001) &10.07713(0.25379) &$ 0.00049$(0.00053)\\
\hline
\end{tabular}
\caption{Simulation parameters and results obtained at $L/a$=6.}
\label{tab:simpara_L6_imp0}
\end{table}

\begin{table}[h]
\begin{tabular}{rccrccccr}
$\beta$ & $\kappa$ & Trajs. & plq. & $\delta\tau$ & Acc. & $g_{\rm SF}^2$ & $M$ \\
\hline
96.0000 & 0.1263270 & 22,500 & 0.979420(0.000002) &0.0056 &0.811(0.004) & 0.06434(0.00004) &$ 0.00001$(0.00001)\\
\hline
48.0000 & 0.1272250 & 18,300 & 0.958843(0.000003) &0.0100 &0.818(0.007) & 0.13247(0.00017) &$ 0.00002$(0.00002)\\
\hline
24.0000 & 0.1291450 & 42,300 & 0.917260(0.000004) &0.0125 &0.804(0.003) & 0.28282(0.00023) &$-0.00004$(0.00002)\\
\hline
12.0000 & 0.1335850 & 68,500 & 0.832266(0.000007) &0.0167 &0.828(0.005) & 0.65380(0.00071) &$-0.00010$(0.00003)\\
\hline
9.6000 & 0.1361800 & 21,820 & 0.788830(0.000013) &0.0200 &0.828(0.006) & 0.88751(0.00287) &$-0.00008$(0.00005)\\
\hline
7.4000 & 0.1406600 & 63,330 & 0.723081(0.000010) &0.0250 &0.818(0.003) & 1.34182(0.00417) &$-0.00004$(0.00016)\\
\hline
6.8000 & 0.1426200 & 41,500 & 0.697409(0.000013) &0.0250 &0.797(0.002) & 1.56232(0.00662) &$ 0.00012$(0.00011)\\
\hline
6.3000 & 0.1447000 & 28,000 & 0.672208(0.000021) &0.0250 &0.816(0.003) & 1.81987(0.01036) &$-0.00034$(0.00014)\\
\hline
6.0000 & 0.1462000 & 47,000 & 0.654999(0.000012) &0.0250 &0.820(0.003) & 2.01248(0.01258) &$-0.00042$(0.00011)\\
\hline
5.5000 & 0.1492700 & 35,900 & 0.622016(0.000021) &0.0286 &0.797(0.003) & 2.48139(0.01969) &$-0.00021$(0.00015)\\
\hline
5.0000 & 0.1533600 & 27,900 & 0.582458(0.000038) &0.0250 &0.825(0.004) & 3.46930(0.07238) &$ 0.00094$(0.00034)\\
\hline
4.8000 & 0.1554270 & 114,500 & 0.564464(0.000020) &0.0250 &0.860(0.001) & 4.35348(0.09845) &$ 0.00026$(0.00024)\\
\hline
4.7000 & 0.1565500 & 35,400 & 0.554789(0.000040) &0.0256 &0.854(0.002) & 4.87595(0.21035) &$ 0.00027$(0.00051)\\
\hline
4.6200 & 0.1575500 & 86,300 & 0.546856(0.000030) &0.0312 &0.783(0.001) & 6.23744(0.25321) &$-0.00023$(0.00033)\\
\hline
4.6000 & 0.1577800 & 149,300 & 0.544695(0.000027) &0.0250 &0.852(0.002) & 6.01108(0.17093) &$-0.00007$(0.00028)\\
\hline
4.5500 & 0.1584200 & 24,500 & 0.539428(0.000090) &0.0278 &0.833(0.003) & 6.92022(0.46491) &$ 0.00087$(0.00088)\\
4.5500 & 0.1584270 & 93,300 & 0.539336(0.000033) &0.0278 &0.831(0.002) & 6.99432(0.31873) &$ 0.00135$(0.00041)\\
4.5500 & 0.1584500 & 25,700 & 0.539683(0.000064) &0.0278 &0.832(0.004) & 6.74187(0.46970) &$-0.00163$(0.00071)\\
\hline
4.5200 & 0.1588500 & 56,570 & 0.536316(0.000058) &0.0278 &0.827(0.002) & 8.28029(0.57687) &$-0.00010$(0.00059)\\
\hline
4.5000 & 0.1591300 & 107,100 & 0.534108(0.000036) &0.0250 &0.859(0.001) & 8.40630(0.37369) &$-0.00007$(0.00038)\\
\hline
4.4800 & 0.1594000 & 41,555 & 0.531781(0.000085) &0.0250 &0.827(0.002) & 8.57214(0.57202) &$ 0.00027$(0.00070)\\
\hline
4.4215 & 0.1602640 & 160,900 & 0.525143(0.000050) &0.0263 &0.837(0.001) &12.21877(0.49625) &$-0.00012$(0.00041)\\
4.4215 & 0.1602700 & 127,500 & 0.525149(0.000058) &0.0250 &0.861(0.001) &12.62365(0.68980) &$-0.00059$(0.00048)\\
\hline
4.4200 & 0.1602700 & 29,700 & 0.524651(0.000132) &0.0278 &0.828(0.002) &13.15085(0.99774) &$ 0.00214$(0.00075)\\
\hline
4.4000 & 0.1606000 & 229,500 & 0.522502(0.000057) &0.0278 &0.819(0.002) &15.00764(0.69115) &$ 0.00020$(0.00042)\\
\hline
\end{tabular}
\caption{Simulation parameters and results obtained at $L/a$=8.}
\label{tab:simpara_L8_imp0}
\end{table}

\begin{table}[h]
\begin{tabular}{rccrccccr}
$\beta$ & $\kappa$ & Trajs. & plq. & $\delta\tau$ & Acc. & $g_{\rm SF}^2$ & $M$ \\
\hline
48.0000 & 0.1269700 & 11,200 & 0.958648(0.000002) &0.0056 &0.815(0.003) & 0.13304(0.00033) &$-0.00014$(0.00003)\\
\hline
24.0000 & 0.1288929 & 54,620 & 0.916777(0.000002) &0.0083 &0.798(0.002) & 0.28432(0.00036) &$-0.00008$(0.00002)\\
\hline
12.0000 & 0.1333359 & 68,955 & 0.831306(0.000003) &0.0125 &0.808(0.002) & 0.66007(0.00119) &$-0.00012$(0.00003)\\
\hline
9.6000 & 0.1359350 & 86,700 & 0.787681(0.000003) &0.0133 &0.806(0.002) & 0.90325(0.00233) &$-0.00001$(0.00003)\\
\hline
7.4000 & 0.1404060 & 106,050 & 0.721824(0.000004) &0.0154 &0.795(0.004) & 1.36896(0.00543) &$-0.00001$(0.00004)\\
\hline
6.8000 & 0.1423250 & 45,150 & 0.696157(0.000006) &0.0167 &0.819(0.002) & 1.59998(0.00983) &$ 0.00091$(0.00006)\\
\hline
6.3000 & 0.1444050 & 23,500 & 0.671015(0.000014) &0.0182 &0.767(0.002) & 1.89012(0.01692) &$ 0.00013$(0.00012)\\
\hline
6.0000 & 0.1459000 & 43,296 & 0.653900(0.000008) &0.0182 &0.820(0.007) & 2.10612(0.02202) &$ 0.00011$(0.00007)\\
\hline
5.8000 & 0.1470200 & 43,400 & 0.641479(0.000007) &0.0182 &0.799(0.002) & 2.22171(0.02802) &$-0.00009$(0.00007)\\
\hline
5.5000 & 0.1489400 & 45,200 & 0.621162(0.000012) &0.0167 &0.842(0.003) & 2.58933(0.02496) &$ 0.00017$(0.00017)\\
\hline
5.2000 & 0.1512000 & 68,000 & 0.598557(0.000009) &0.0200 &0.781(0.002) & 3.06212(0.03210) &$ 0.00056$(0.00009)\\
\hline
5.0800 & 0.1522350 & 25,480 & 0.588819(0.000013) &0.0167 &0.848(0.002) & 3.39969(0.08924) &$ 0.00015$(0.00013)\\
\hline
5.0000 & 0.1529700 & 90,921 & 0.582135(0.000019) &0.0167 &0.826(0.002) & 3.67356(0.08908) &$-0.00044$(0.00020)\\
\hline
4.8000 & 0.1549700 & 194,300 & 0.564213(0.000006) &0.0182 &0.810(0.002)& 4.84805(0.14509) &$ 0.00009$(0.00011)\\
\hline
4.6000 & 0.1572300 & 127,922 & 0.544423(0.000012) &0.0182 &0.818(0.002) & 7.29885(0.38589) &$ 0.00075$(0.00017)\\
\hline
4.5500 & 0.1578500 & 57,260 & 0.539025(0.000021) &0.0192 &0.802(0.002)  &10.15231(1.11827) &$ 0.00125$(0.00027)\\
\hline
4.5000 & 0.1585500 & 104,570 & 0.534014(0.000049) &0.0250 &0.701(0.007) &13.03915(1.33994) &$-0.00237$(0.00039)\\
\hline

\end{tabular}
\caption{Simulation parameters and results obtained at $L/a$=12.}
\label{tab:simpara_L12_imp0}
\end{table}

\begin{table}[h]
\begin{tabular}{rccrccccr}
$\beta$ & $\kappa$ & Trajs. & plq. & $\delta\tau$ & Acc. & $g_{\rm SF}^2$ & $M$ \\
\hline
24.0000 & 0.1288000 & 22,890 & 0.916505(0.000002) &0.0067 &0.796(0.003) & 0.28368(0.00057) &$ 0.00011$(0.00002)\\
\hline
12.0000 & 0.1332590 & 40,950 & 0.830799(0.000003) &0.0091 &0.718(0.011) & 0.66384(0.00293) &$-0.00016$(0.00002)\\
\hline
9.6000 & 0.1358600 & 30,900 & 0.787096(0.000002) &0.0080 &0.807(0.003) & 0.90538(0.00570) &$-0.00023$(0.00003)\\
\hline
7.4000 & 0.1403250 & 62,300 & 0.721190(0.000003) &0.0111 &0.812(0.002) & 1.39094(0.00834) &$-0.00002$(0.00003)\\
\hline
6.8000 & 0.1422900 & 39,796 & 0.695613(0.000004) &0.0133 &0.787(0.002) & 1.63562(0.01649) &$-0.00042$(0.00005)\\
\hline
6.3000 & 0.1443400 & 69,000 & 0.670503(0.000004) &0.0133 &0.798(0.002) & 1.91412(0.01628) &$-0.00036$(0.00004)\\
\hline
6.0000 & 0.1457950 & 18,900 & 0.653363(0.000007) &0.0156 &0.712(0.004) & 2.12147(0.03887) &$ 0.00043$(0.00010)\\
\hline
5.5000 & 0.1488500 & 50,330 & 0.620911(0.000007) &0.0143 &0.782(0.002) & 2.67936(0.03897) &$-0.00047$(0.00011)\\
\hline
5.0800 & 0.1521310 & 23,760 & 0.588803(0.000007) &0.0139 &0.804(0.003) & 3.24742(0.07271) &$-0.00003$(0.00012)\\
\hline
5.0000 & 0.1528550 & 71,954 & 0.582121(0.000004) &0.0143 &0.797(0.002) & 3.86709(0.12622) &$-0.00004$(0.00009)\\
\hline
4.8000 & 0.1548310 & 46,000 & 0.564445(0.000008) &0.0156 &0.755(0.002) & 5.72911(0.49013) &$ 0.00003$(0.00014)\\
\hline
4.6000 & 0.1570500 & 83,705 & 0.544764(0.000008) &0.0143 &0.794(0.001) & 8.21243(0.63114) &$ 0.00128$(0.00012)\\
\hline
4.5500 & 0.1576750 & 107,069 & 0.539609(0.000010) &0.0139 &0.809(0.002) &10.81452(0.80073) &$ 0.00011$(0.00011)\\
\hline
4.5200 & 0.1580650 & 42,400 & 0.536387(0.000021) &0.0156 &0.754(0.002) &17.34193(3.72829) &$-0.00030$(0.00021)\\
\hline
\end{tabular}
\caption{Simulation parameters and results obtained at $L/a$=16.}
\label{tab:simpara_L16_imp0}
\end{table}

\begin{table}[h]
\begin{tabular}{rccrccccr}
$\beta$ & $\kappa$ & Trajs. & plq. & $\delta\tau$ & Acc. & $g_{\rm SF}^2$ & $M$ \\
\hline
4.5500 & 0.1576500 & 32,309 & 0.540093(0.000014) &0.0143 &0.785(0.003) &11.13131(1.41381) &$-0.00124$(0.00018)\\
\hline
\end{tabular}
\caption{Simulation parameters and results obtained at $L/a$=18.}
\label{tab:simpara_L18_imp0}
\end{table}

\begin{table}[htb]
\centering
\begin{tabular}{cc|cccccc}
$L/a$ & $N$ & $\chi^2$/dof & $a_{L/a,1}$ & $a_{L/a,2}$ & $a_{L/a,3}$ & $a_{L/a,4}$ & $a_{L/a,5}$ \\
\hline
  6 &  3  & 9.0(1.3) &  0.4906( 0.0025) & -0.2749( 0.0105) & -0.1897( 0.0151)\\
  6 &  4  & 1.4(0.5) &  0.5048( 0.0014) & -0.3993( 0.0119) &  0.1136( 0.0283) & -0.2042( 0.0184)\\
  6 &  5  & 1.3(0.5) &  0.5015( 0.0032) & -0.4240( 0.0256) &  0.2538( 0.1301) & -0.4043( 0.1815) &  0.0899( 0.0808)\\
\hline
  8 &  3  & 2.1(0.7) &  0.5068( 0.0018) & -0.2308( 0.0104) & -0.2412( 0.0150)\\
  8 &  4  & 0.6(0.3) &  0.5153( 0.0019) & -0.3410( 0.0260) &  0.0405( 0.0629) & -0.1852( 0.0390)\\
  8 &  5  & 0.6(0.8) &  0.5153( 0.0051) & -0.3419( 0.1697) &  0.0444( 0.7500) & -0.1904( 0.9672) &  0.0021( 0.3906)\\
\hline
 12 &  3  & 3.0(1.0) &  0.5239( 0.0047) & -0.1923( 0.0118) & -0.3019( 0.0198)\\
 12 &  4  & 1.1(0.6) &  0.5400( 0.0038) & -0.3614( 0.0376) &  0.1063( 0.0884) & -0.2671( 0.0550)\\
 12 &  5  & 1.0(0.6) &  0.5438( 0.0039) & -0.2783( 0.0726) & -0.2779( 0.2977) &  0.2457( 0.3815) & -0.2165( 0.1582)\\
\hline
 16 &  3  & 4.9(1.4) &  0.5308( 0.0055) & -0.1881( 0.0266) & -0.3057( 0.0375)\\
 16 &  4  & 1.8(0.8) &  0.5520( 0.0039) & -0.4948( 0.0663) &  0.4387( 0.1516) & -0.4762( 0.0903)\\
 16 &  5  & 1.9(0.9) &  0.5538( 0.0050) & -0.4403( 0.1324) &  0.1801( 0.5648) & -0.1283( 0.7332) & -0.1457( 0.3025)\\
\\
\end{tabular}
\caption{The results for the coefficients in the fit function
(\ref{eq:fitfunc})
}
\label{tab:fit_param1_pade_4}
\end{table}


Hereafter we denote the SF coupling obtained at a bare coupling
constant $g_0^2$ and at a lattice length of $l$ by
$g_{\rm SF}^2(g_0^2,l)$ and its continuum counterpart by
$g_{\rm SF}^2(L)$.

\subsection{Discrete $\beta$ function}
\label{subsec:DBF}

In order to see the scale dependence of the SF coupling constant, we
analyze the discrete $\beta$ function (DBF) introduced in
Refs.~\cite{Shamir:2008pb,DeGrand:2010na}.
The whole procedure is described below.

First, we choose an initial value of the running coupling constant,
denoted by $u$.
This implicitly sets the initial length scale $L_0$ through
$g_{\rm SF}^2(L_0)=u$.
Using the interpolating formula (\ref{eq:fitfunc}) for the lattice size
$l$ ($=L/a$), the bare coupling constant $g_0^*$ is numerically obtained
by solving the equation $g_{\rm SF}^2({g_0^*}^2,\,l)=u$.
$l$ is identified with $L_0/a$, so that the lattice spacing at
${g_0^*}^2$ is found to be $a({g_0^*}^2,l)=L_0/l$.
Now we choose a rescaling factor, $s$.
The lattice step scaling function $\Sigma_0(u,s,l)$ is then defined as
the SF coupling for $l'=s\cdot l$ at the same bare coupling
${g_0^*}^2$, {\it i.e.}
\begin{eqnarray}
\Sigma_0(u,s,l) \equiv
  \left. g_{\rm SF}^2({g_0^*}^2, s \cdot l)
  \right|_{ g_{\rm SF}^2({g_0^*}^2,l) = u }.
\end{eqnarray}
The meaning of the subscript ``0'' becomes clear soon.
Of course, both $l$ and $s \cdot l$ must be equal to one of 6, 8, 12
and 16, and hence the possible values for the rescaling factor $s$
are limited.
The difference between $\Sigma_0(u,s,l)$ and $u$ gives the scale
dependence through the scale change from $L$ to $s\cdot L$, up to
lattice artifacts.

Since the raw data of $1/g_{\rm SF}^2(g_0^2, l)$ fluctuate around zero
in the strong coupling region, converting from
$1/g_{\rm SF}^2(g_0^2, l)$ to $g_{\rm SF}^2(g_0^2,l)$ sometimes induces
huge statistical uncertainty.
To avoid this we treat the inverse coupling constant,
$1/g_{\rm SF}^2(g_0^2,l)$, directly.
Then, to see the scale dependence of the inverse coupling constant, we
introduce the lattice DBF~\cite{Shamir:2008pb,DeGrand:2010na} by
\begin{eqnarray}
   B_0(u,s,l)
 = \frac{1}{\Sigma_0(u,s,l)} - \frac{1}{u}.
 \label{eq:dbf-0}
\end{eqnarray}
We calculate the continuum limit of this function for various initial
values of the coupling constant, $u$.
If the sign of the DBF in the continuum limit turns out to flip at a
certain renormalized coupling constant $u$, it indicates the existence
of IRFP.

\subsection{improving discretization errors}
\label{subsec:improving-dbf}

Since $O(a)$ discretization errors are not improved at all in the
lattice actions, it is important to remove the scaling violation as much
as possible.
To do this, we perform the following improvements on the step scaling
function and the DBF before taking the continuum limit.

First let $\sigma(u,s)$ be the continuum limit of $\Sigma_0(u,s,l)$,
{\it i.e.} $\sigma(u,s)=g_{\rm SF}^{2}(s L)$ with
$u=g_{\rm SF}^{2}(L)$.
Its perturbative expression is given by
\begin{eqnarray}
&&  \sigma(u,s) = u + s_0 u^2 + s_1 u^3 + s_2 u^4 
       + \cdots,
       \label{eq:sigma}\\
&& s_0 = {b_1} \ln (s),\\
&& s_1 = \ln (s) \left({b_1}^2 \ln (s)+{b_2}\right),\\
&& s_2 = \ln(s)\left(
                  {b_1}^3 \ln^2(s)
                + \frac{5}{2} {b_1} {b_2} \ln(s)
                + {b_3}
               \right),
\end{eqnarray}
where $b_i$'s are the coefficients of the $\beta$-function introduced in
sec.~\ref{sec:ptanalysis}.
Recalling the parametric form of the discretization
error~\cite{Bode:1999sm}, the error normalized by $\sigma(u,s)$, denoted
by $\delta_0(u,s,l)$, is written as
\begin{eqnarray}
   \delta_0(u,s,l)
 = \frac{\Sigma_0(u,s,l)-\sigma(u,s)}
        {\sigma(u,s)}
  = \delta^{(1)}(s,l)\, u + \delta^{(2)}(s,l)\, u^{2} + O(u^3).
 \label{eq:Sig-sig-0}
\end{eqnarray}
With eq.~(\ref{eq:sigma}), the discretization error at the lowest order
in $u$ is found to be
\begin{eqnarray}
   \delta^{(1)}(s,l)
 = \Big( p_{1,s \cdot l}-b_{1}\ln(s \cdot l) \Big)
 - \Big( p_{1,        l}-b_{1}\ln(        l) \Big)
 = p_{1,s \cdot l}-p_{1,l}-b_1\ln(s).
 \label{eq:delta_1}
\end{eqnarray}
Now by replacing $\Sigma_0(u,s,l)$ in eq.~(\ref{eq:Sig-sig-0}) with
$\Sigma_1(u,s,l)=\Sigma_0(u,s,l)/(1+\delta^{(1)}(s,l)\,u)$, the
discretization error reduces to $O(u^2)$.
Using $\Sigma_1(u,s,l)$, the one-loop improved DBF is defined by
\begin{eqnarray}
     B_1(u,s,l)
 &=& \frac{1}{\Sigma_1(u,s,l)} - \frac{1}{u}.
 \label{eq:dbf-1}
\end{eqnarray}
This completes the one-loop improvement.

The above procedure can be repeated to an arbitrarily higher order in
$u$, but it requires the perturbative coefficients like $p_{1,l}$ and
the perturbative expression of $\sigma(u,s)$ to the corresponding order
in $u$.
All the coefficients necessary for the two-loop improvement are not
available at this moment.
Instead, we follow an alternative prescription proposed in
Ref.~\cite{Aoki:2009tf}.
After the one-loop improvement, the scaling violation is written as
\begin{eqnarray}
   \delta_1(u,s,l)
 = \frac{\Sigma_1(u,s,l) - \sigma(u,s)}
        {\sigma(u,s)}
  = \delta^{(2)}(s,l)\, u^{2}+ O(u^3).
 \label{eq:Sig-sig}
\end{eqnarray}
If one can somehow know $\delta^{(2)}(s,l)$, the scaling violation
can be reduced to $O(u^3)$ by replacing $\Sigma_0(u,s,l)$
in eq.~(\ref{eq:Sig-sig-0}) with
\begin{eqnarray}
    \Sigma_2(u,s,l)
  = \Sigma_0(u,s,l)/(1+\delta^{(1)}(s,l)\,u+\delta^{(2)}(s,l)\,u^2).
\label{eq:Sig-sig-2}
\end{eqnarray}
$\delta^{(2)}(s,l)$ can be determined by fitting our data for
$\delta_1(u,s,l)$ in eq.~(\ref{eq:Sig-sig}) to the function quadratic in
$u$.
Notice that in order for this fitting to make sense, the perturbative
series of $\sigma(u,s)$ must be known through $O(u^3)$.
Since the first two coefficients, $b_{1}$ and $b_{2}$, are available,
the correct value of $\sigma(u,s)$ can be calculated to $O(u^{3})$ as
seen from eq.~(\ref{eq:sigma}).

\begin{figure}[tb]
\centering
\begin{tabular}{cc}
\includegraphics*[width=0.5 \textwidth,clip=true]
{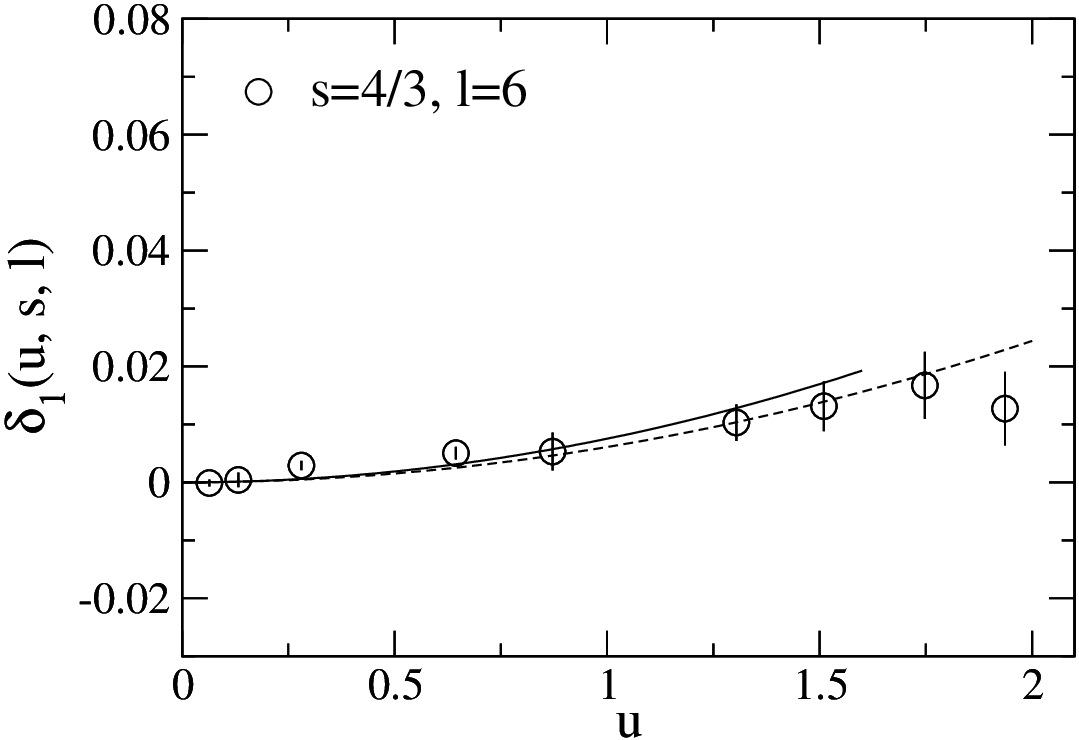}&
\includegraphics*[width=0.5 \textwidth,clip=true]
{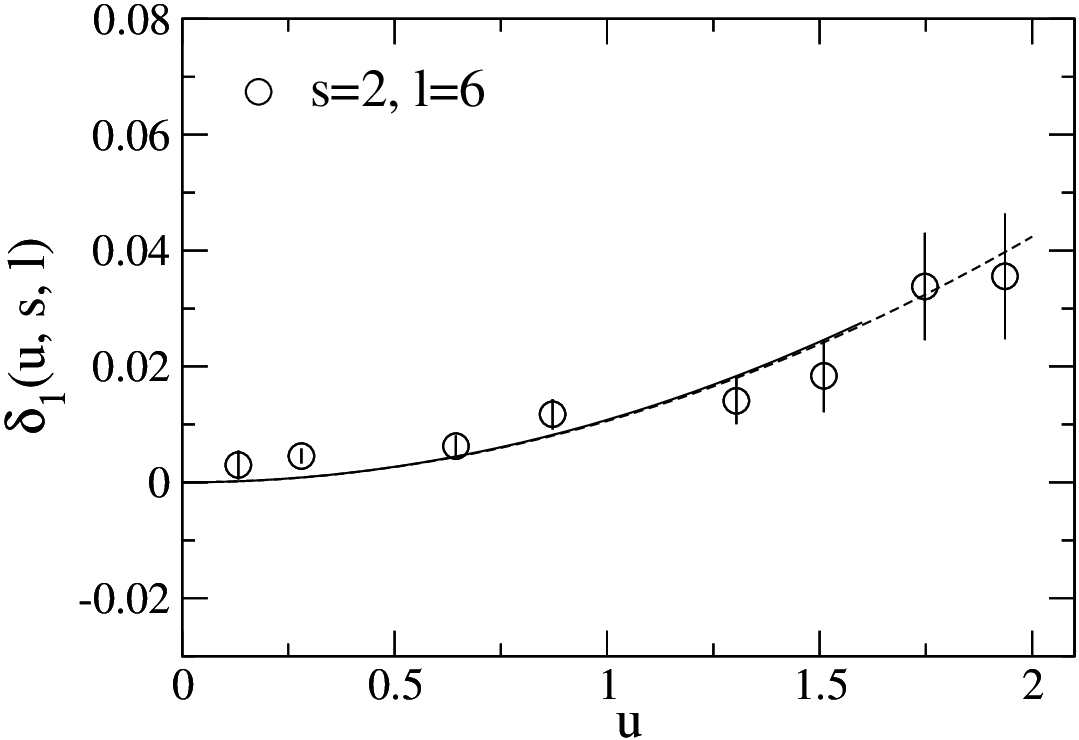}\\
\includegraphics*[width=0.5 \textwidth,clip=true]
{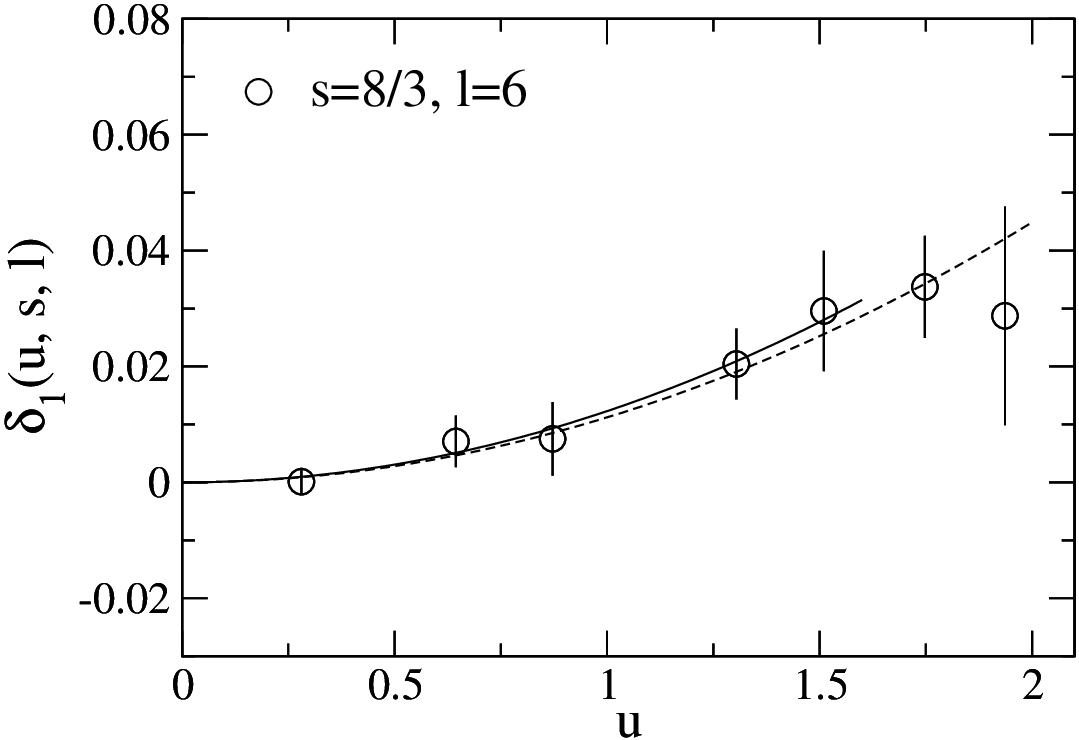}&
\includegraphics*[width=0.5 \textwidth,clip=true]
{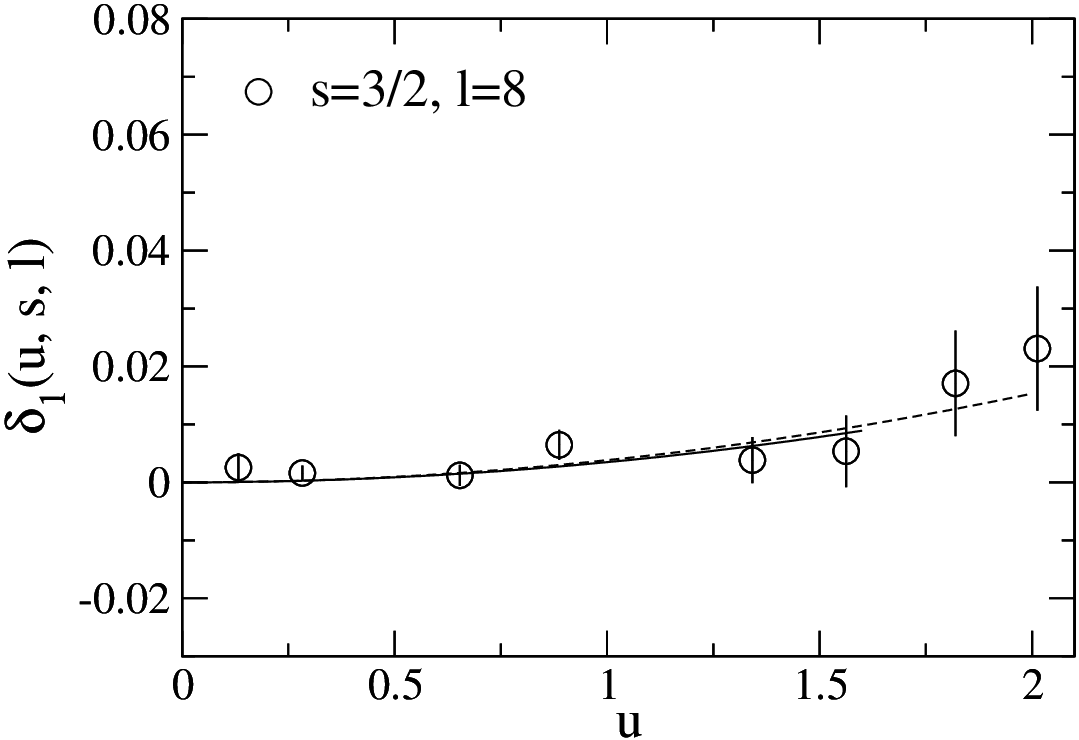}\\
\includegraphics*[width=0.5 \textwidth,clip=true]
{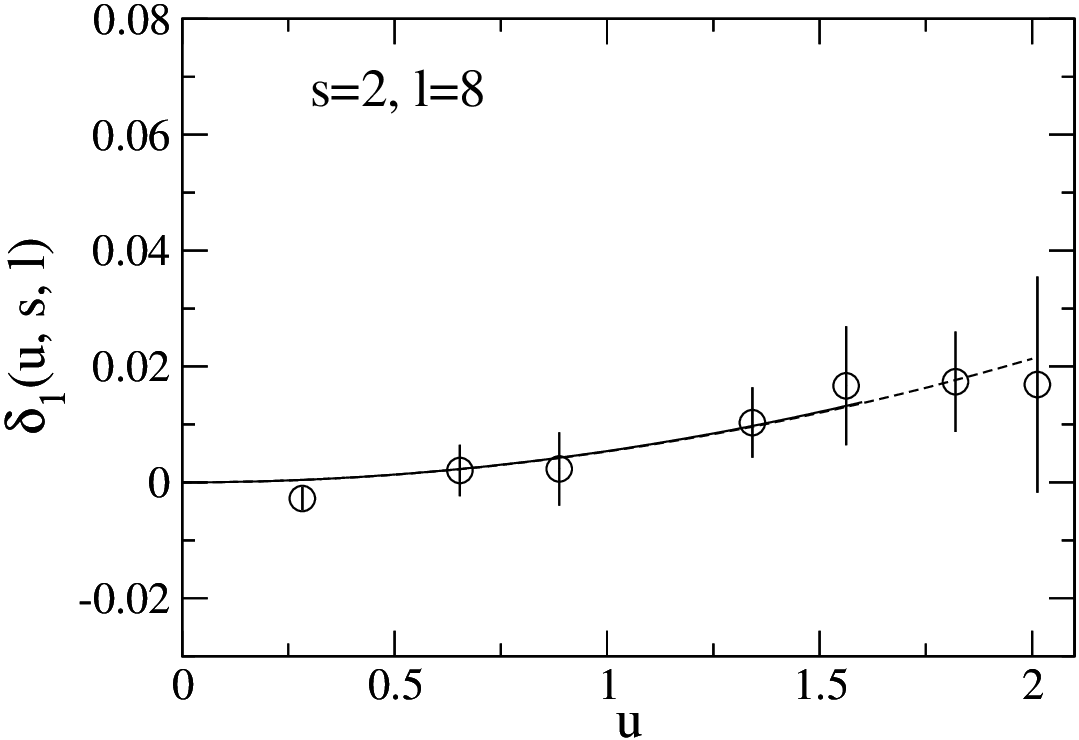}&
\includegraphics*[width=0.5 \textwidth,clip=true]
{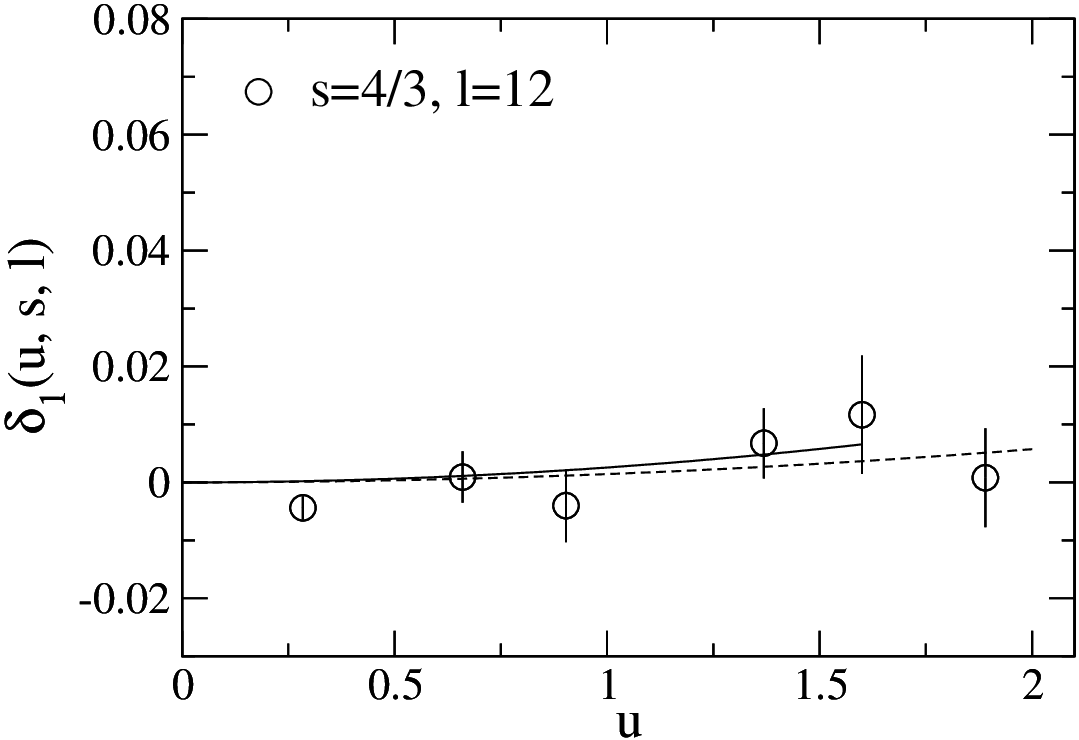}
\end{tabular}
\caption{Fit of $\delta_1$ to a quadratic function of $u$.
 The solid and dashed curves show the fit results and the fit ranges.
 }
\label{fig:deltas}
\end{figure}

$\delta_1(u,s,l)$ is fitted to the form of eq.~(\ref{eq:Sig-sig}),
neglecting $O(u^3)$ or higher order terms,
for all possible pairs of $(s,l)$ as shown in Fig.~\ref{fig:deltas}.
The fit has to be performed in a weak coupling region where the
perturbative expansion is reliable.
We examine two fit ranges, $0 \le u \le 1.6$ and $0 \le u \le 2.0$ to
see the fit range dependence.
The extracted values for $\delta^{(2)}(s,l)$ are tabulated in
Tab.~\ref{tab:delta2} together with $\delta^{(1)}(s,l)$ defined in
eq.~(\ref{eq:delta_1}).

The table shows that the values of $\delta^{(1)}(s,l)$ and
$\delta^{(2)}(s,l)$ lie between $10^{-2}$ and $10^{-3}$, and
$\delta^{(2)}(s,l)$ turns out not to depend on the fit range.
In the following analysis, we employ $\delta^{(2)}(s,l)$ from the
shorter fit range.
It is also seen from the table that generally the coefficients for
$(s, l)=(4/3, 12)$ are the smallest among others.
This is anticipated because the improvement coefficient vanish as $s$
approaches to unity or $l$ becomes large.
An exception is the one-loop coefficient $\delta^{(1)}(4/3,6)$.
Since two-loop coefficient $\delta^{(2)}(4/3,6)$ is, however, much
larger than $\delta^{(1)}(4/3,6)$, the smallness of
$\delta^{(1)}(4/3,6)$ is probably by accident.
In the data sets we have, the data with $(s,\,l)=(4/3,\,6)$ is the
coarsest one.
As we will show in the following subsections, this data turns out to
suffer from non-linear scaling violation larger than the linear one in
the strong coupling region.
Thus, we omit this data point throughout the analysis.
Using $\delta^{(2)}(s,l)$ thus obtained, we define the two-loop improved
step scaling function $\Sigma_2(u,s,l)$ in eq.~(\ref{eq:Sig-sig-2}), and
in turn the two-loop improved DBF
\begin{eqnarray}
     B_2(u,s,l)
 &=& \frac{1}{\Sigma_2(u,s,l)} - \frac{1}{u}\,.
 \label{eq:dbf-2}
\end{eqnarray}
\begin{table}
 \begin{tabular}{c|cccccc}
  $(s,l)$ & $(4/3, 6)$ & $(2, 6)$ & $(8/3, 6)$ & $(3/2, 8)$ & $(2, 8)$ & $(4/3, 12)$
  \\
\hline\hline
  $\delta^{(1)}(s,l)$ & $-$0.00102   & $-$0.0101    & $-$0.0182
                      & $-$0.00905   & $-$0.0172    & $-$0.00817 \\
\hline\hline
  $\delta^{(2)}(s,l)$ [0, 1.60]
                      & 0.0075(12) & 0.0108(15) & 0.0123(26)
                      & 0.0035(14) & 0.0054(24) & 0.0026(23)\\
  $\chi^2$/dof & 2.2 & 3.0 & 0.1 & 1.0 & 0.7 & 1.7 \\
\hline
  $\delta^{(2)}(s,l)$ [0, 2.0]
                      & 0.0061(9)  & 0.0106(12) & 0.0112(18)
                      & 0.0038(13) & 0.0053(18) & 0.0014(17)\\
  $\chi^2$/dof & 2.2 & 2.1 & 0.2 & 0.9 & 0.6 & 1.4 \\
 \end{tabular}
 \caption{Coefficients for perturbative correction, $\delta^{(1)}(s,l)$
 and $\delta^{(2)}(s,l)$, for each pair of $(s,l)$.
 The square brackets in the first column indicate the fit range in $u$.
 }
 \label{tab:delta2}
\end{table}

\subsection{strategy}
The continuum limit is taken for a fixed rescaling factor $s$ and a
fixed input length scale $L$ varying a lattice spacing $a$ $(=L/l)$.
As described in the preceding subsections, an input length scale is
fixed by choosing a particular value of input coupling $u$.
However, for a given $s$ the number of data sets with different
$a$ in this work is, at most, two; $(s,l)=(2,6)$ and (2,8) for $s=2$.
While it is still possible to employ these two sets of data to evaluate
the continuum limit, the validity of the linear extrapolation can not be
tested.
Alternatively, we may supplement a data set with a desired $s$
by interpolating or extrapolating data of $g_{\rm SF}^2(g_0^2,l)$ in
$l$.
However, a lack of guiding principles in the interpolation or
extrapolation may cause a systematic uncertainty.
In this work, we use the two available data sets, $(s,l)=(2,6)$ and
(2,8) to evaluate the continuum limit by linear extrapolation, and the
other data sets are used to monitor the validity of the linear
extrapolation.
For this purpose,
we introduce a relation which approximately converts the DBF for $s'$
into that for $s$, as follows.

We start with a closer look at the discretization error.
The discretization error of the lattice DBF, {\it i.e.}
$B_i(u,s,l)- B(u,s)$ ($i$=0, 1, 2), can be expressed
in terms of an asymptotic expansion in $1/l$~\cite{Bode:1999sm} as
\begin{eqnarray}
    B_i(u,s,l) - B(u,s)
&=& \left(\frac{1}{l}-\frac{1}{s\,l}\right) e_i(u) + O(l^{-2}),
\label{eq:error-i}
\end{eqnarray}
where $e_i(u)$ is an unknown coefficient of $O(a)$ error and is a
function of $u$.
We then define the {\em rescaled} lattice DBF by
\begin{eqnarray}
   B_i'(u,s,l,s')&=&\frac{\ln(s)}{\ln(s')}B_i(u,s',l).
   \label{eq:rescale-lat-DBF}
\end{eqnarray}
In addition, using the continuum counterpart of
eq.~(\ref{eq:rescale-lat-DBF}), we define
\begin{eqnarray}
   \delta B(u,s,s')
 = B(u,s) - \frac{\ln(s)}{\ln(s')}\,B(u,s'),
 \label{eq:delta-B}
\end{eqnarray}
which represents the difference between the true continuum DBF and the
{\it rescaled} continuum DBF.
Combining eqs.~(\ref{eq:error-i}), (\ref{eq:rescale-lat-DBF}) and
(\ref{eq:delta-B}) together and introducing
\begin{eqnarray}
   \xi(s,l,s')
 = \frac{\ln(s)}{\ln(s')} \left(\frac{1}{l} - \frac{1}{s' l}\right),
  \label{eq:def_xi}
\end{eqnarray}
we arrive at
\begin{eqnarray}
   B_i'(u,s,l,s')
 = B(u,s) + \xi(s,l,s')\ e_i(u) - \delta B(u,s,s') + O(l^{-2}).
\label{eq:error-2}
\end{eqnarray}
Therefore, if $\delta B(u,s,s')$ and $O(l^{-2})$ (or higher order)
discretization errors are negligible compared to the statistical error
of $B_i'(u,s,l,s')$, the numerical data of $B_i'(u,s,l,s')$ plotted
against $\xi$ will line up on a single line, and even two unknown
coefficients in eq.~(\ref{eq:error-2}), $B(u,s)$ and $e_i(u)$, for given
$u$ and $s$ can be extracted from that behavior.
Instead, if both or one of them is large, the data will not align.
Thus, whether $B_i'(u,s,l,s')$ plotted against $\xi$ aligns or not tests
the validity of the linear extrapolation within the statistical
uncertainty.

We comment on the size of $\delta B(u,s,s')$.
Solving eq.~(\ref{eq:betafunc}) perturbatively, the continuum DBF is
found to be
\begin{eqnarray}
     B(u,s)
 &=& - \ln(s)\,\Bigg\{
       \frac{\beta(u)}{u^2}
     + u^2 \ln(s) \frac{1}{2} {b_1} {b_2}
     + u^3 \ln(s)
       \left(  \frac{1}{3} {b_1}^2 {b_2} \ln(s)
              + {b_1} {b_3} 
              + \frac{1}{2} {b_2}^2
           \right)
     \Bigg\}
\no\\&&
     +O\left( u^4 \ln^2(s) \right),
\end{eqnarray}
and thus the perturbative expression of $\delta B(u,s,s')$ is
\begin{eqnarray}
      \delta B(u,s,s')
 &=& u^2\ln(s) \ln\left(\frac{s}{s'}\right)
     \Bigg[
     - \frac{1}{2} {b_1} {b_2}
     + u \Big\{
        -\frac{1}{3} {b_1}^2 {b_2} \ln(s s')
        -\bigg({b_1} {b_3}+\frac{1}{2} {b_2}^2 \bigg)
            \Big\}
          \Bigg]
\no\\&&
       +O\left(u^4\,\ln(s)\,\ln(s/s')\right).
\end{eqnarray}
Since the numerical values of $b_i$'s are small, {\it e.g.}
$b_1 \sim 0.055$, $b_2\sim -0.002$, $b_3^{\rm SF}\sim O(10^{-4})$,
$\delta B(u,s,s')$ is also small in the perturbative regime as
$10^{-5} \times u^2\,( 1.5 + 0.6\, u)$,
$10^{-5} \times u^2\,( 1.1 + 0.4\, u)$ and
$10^{-5} \times u^2\,(-1.1 - 0.5\, u)$
for $(s,\,s')=(2,\,16/12)$, $(2,\,12/8)$ and $(s,\,s')=(2,\,16/6)$,
respectively.
As $u$ becomes large, $\delta B(u,s,s')$ may become sizable and at some
point exceed the statistical error of $B_i'(u,s,l,s')$.
Then, the alignment will be deformed.
Notice that, the smaller $B(u,s')$ is, the smaller $\delta B(u,s,s')$
is, and in particular, when $B(u,s')=0$ for a certain $s'$, $\delta
B(u,s,s')=0$ holds exactly.

We extract the continuum DBF $B(u,s)$ as follows.
First, we assume linear scaling violation and calculate $B(u,s)$ for
$s$=2 by extrapolating the two data sets, $(s',l)=(2,6)$ and (2,8), to
$\xi=0$.
Since $s'=s$, $B_i'(u,s,l,s')=B_i(u,s',l)$ and $ \delta B(u,s,s')=0$ by
construction.
Thus we do not have to rely on the smallness of $\delta B(u,s,s')$.
Then, to test the linearity of the scaling violation, we calculate
the rescaled lattice DBF $B_i'(u,s,l,s')$ with $s$=2 from the other data
sets and plot them as a function of $\xi(s,l,s')$.
If the data align within the statistical error of $B_i'(u,2,l,s')$, the
assumption of the linear scaling violation is valid, $\delta B(u,s,s')$
is negligible and then the value of $B(u,s)$ thus obtained is reliable.
Alternatively, once the linearity is confirmed, we can even determine
the continuum limit by taking the linear extrapolation of
$B_i'(u,2,l,s')$.
Since $\delta B(u,s,s')$ is negligible in perturbative regime, the
linearity can be tested more rigorously in such a regime.
When the data do not align, either or both of the linear violation
dominance and small $\delta B(u,s,s')$ are invalid and the result for
$B(u,s)$ becomes uncertain.

\subsection{extraction of the continuum DBF}
\label{subsec:contlim}

Extrapolation to the continuum limit described in the following is
carried out for every jack-knife ensemble, and the statistical error in
the continuum limit is estimated by the single elimination jack-knife
method.

\begin{figure}[tb]
\centering
\begin{tabular}{cc}
\includegraphics*[width=0.5 \textwidth,clip=true]
{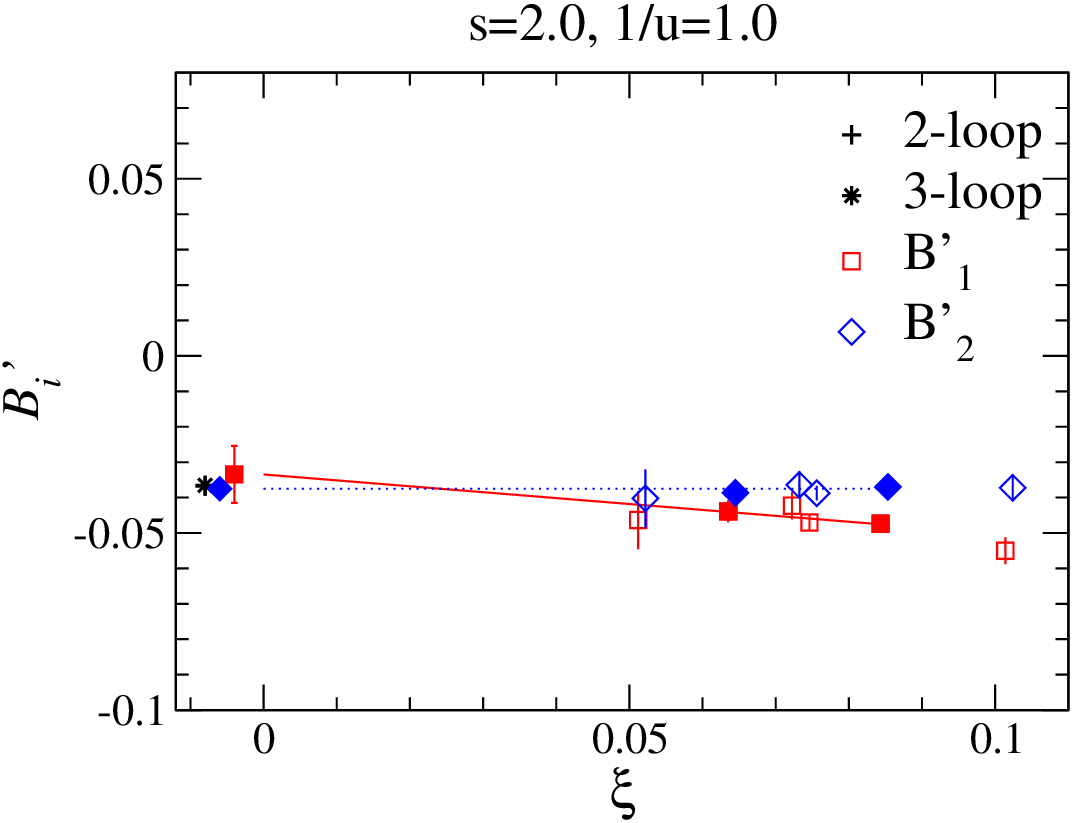}&
\includegraphics*[width=0.5 \textwidth,clip=true]
{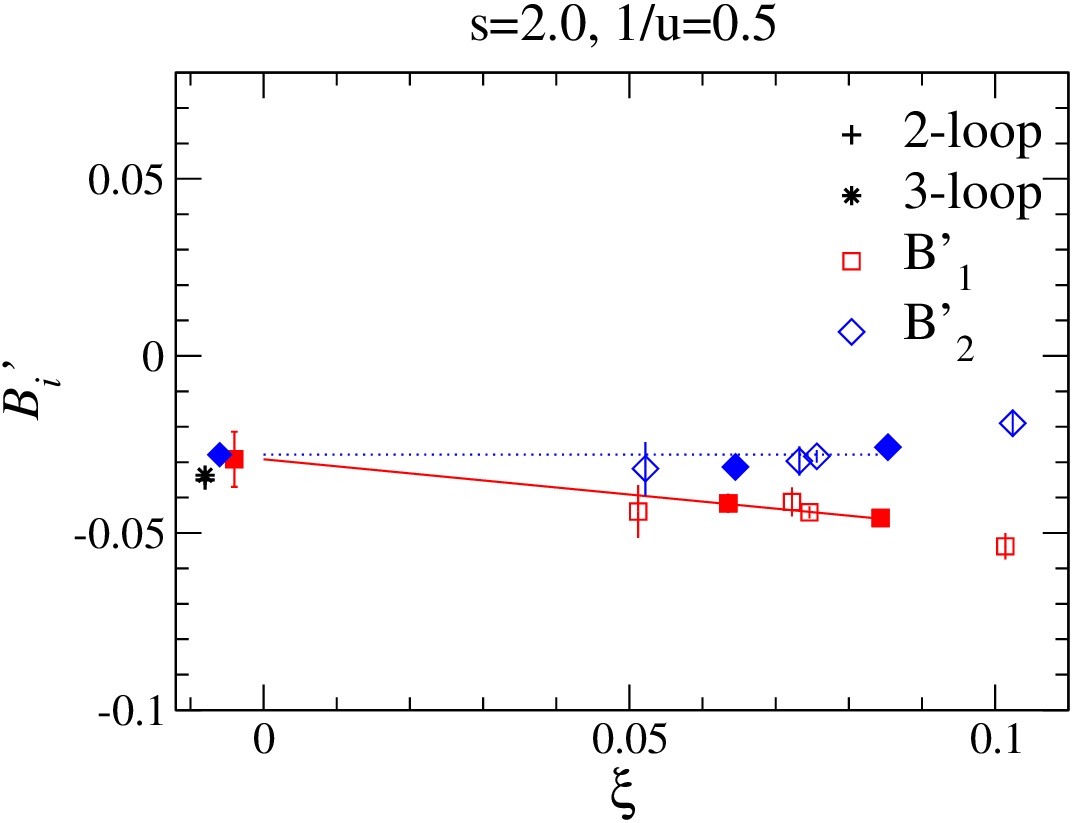}\\
\includegraphics*[width=0.5 \textwidth,clip=true]
{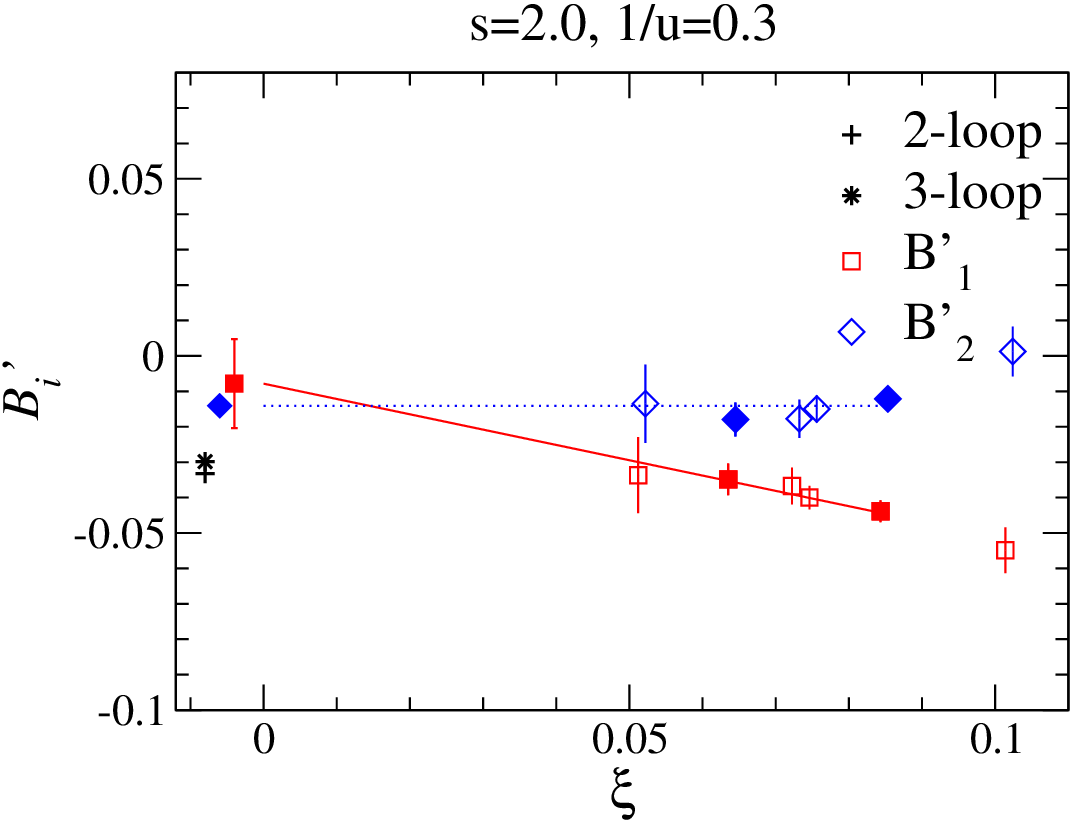}&
\includegraphics*[width=0.5 \textwidth,clip=true]
{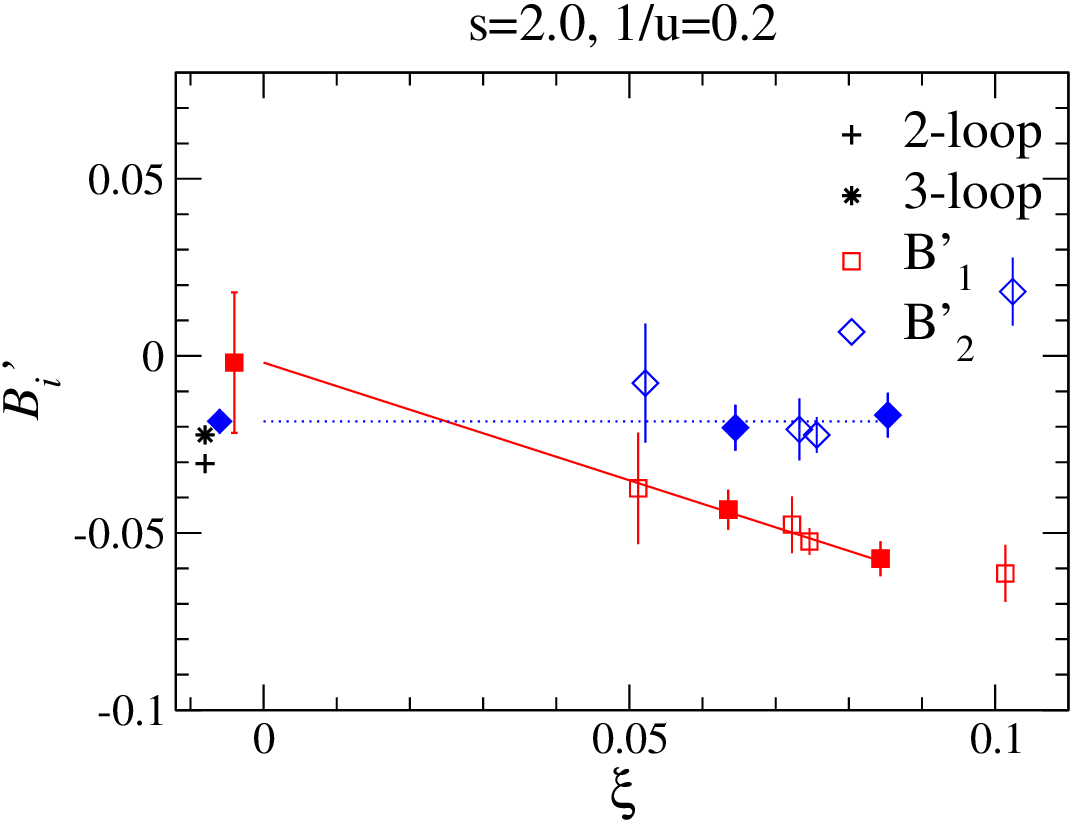}\\
\end{tabular}
\caption{
 Linear extrapolation of $B_1$ (filled squares) and constant fit of
 $B_2'$ (filled diamonds) to the continuum limit.
 The extrapolation and fit use the data with $s'=2$ (filled symbols).
 The data with $s'\ne 2$ (open symbols) are also shown to see whether
 they align or not.
 The values of ($s'$, $l$) of the data shown are (4/3, 6), (2, 6),
 (8/3, 6),  (3/2, 8), (2, 8), and (4/3, 12) from right to left.
 The data points are slightly shifted in horizontal direction for
 clarity.
 The perturbative predictions including the 2-loop (plus) and 3-loop
 (star) effects are also shown.}
\label{fig:contlimit-g2run-each-weak}
\end{figure}

We begin with analysis at relatively weak coupling.
Figure~\ref{fig:contlimit-g2run-each-weak} shows the continuum limit of
$B_i'(u,s,l,s')$ for $s=2$ ($i$=1, 2) at the four representative values
of $1/u$ corresponding to $u=1.0, 2.0, 10/3, 5.0$,
where the data with $s'=2$ are shown in filled symbols and the other in
open symbols and two data of the one-loop improved lattice DBF ($B_1'$)
with $s'=2$ (filled squares) are linearly extrapolated to $\xi=0$.
The two-loop improvement described in sec.~\ref{subsec:improving-dbf} is
equivalent to tuning the improvement coefficients such that the
resulting DBF reproduces the perturbative DBF in the region $0<u<1.6$.
Indeed, the constant fit of the two-loop improved DBF with $s'=2$
(filled diamonds) gives the value consistent with the perturbative
prediction when $1/u=1.0$ and $0.5$ as seen in the figure.
In the same region ($1/u \simg 0.5$), the data of the one-loop improved
DBF align and the linear extrapolation reproduces the perturbative DBF
as well.
Importantly, the extracted continuum DBF is clearly negative in this
region.

The deviation from the perturbative prediction appears at $1/u=0.3$,
where the linear extrapolation gives the value closer to and consistent
with zero.
It is important to note that the data of the one-loop improved DBF align
down to $1/u=0.2$ with a slope increasing with $u$.
From this observation, we conclude that, in the region $1/u \ge 0.2$
($u \le 5$), $\delta B(u,s,s')$ is small, the scaling violation is
linear for the one-loop improved DBF and hence the extracted continuum
limit is reliable.

\begin{figure}[tb]
\centering
\begin{tabular}{c}
\includegraphics*[width=0.5 \textwidth,clip=true]
{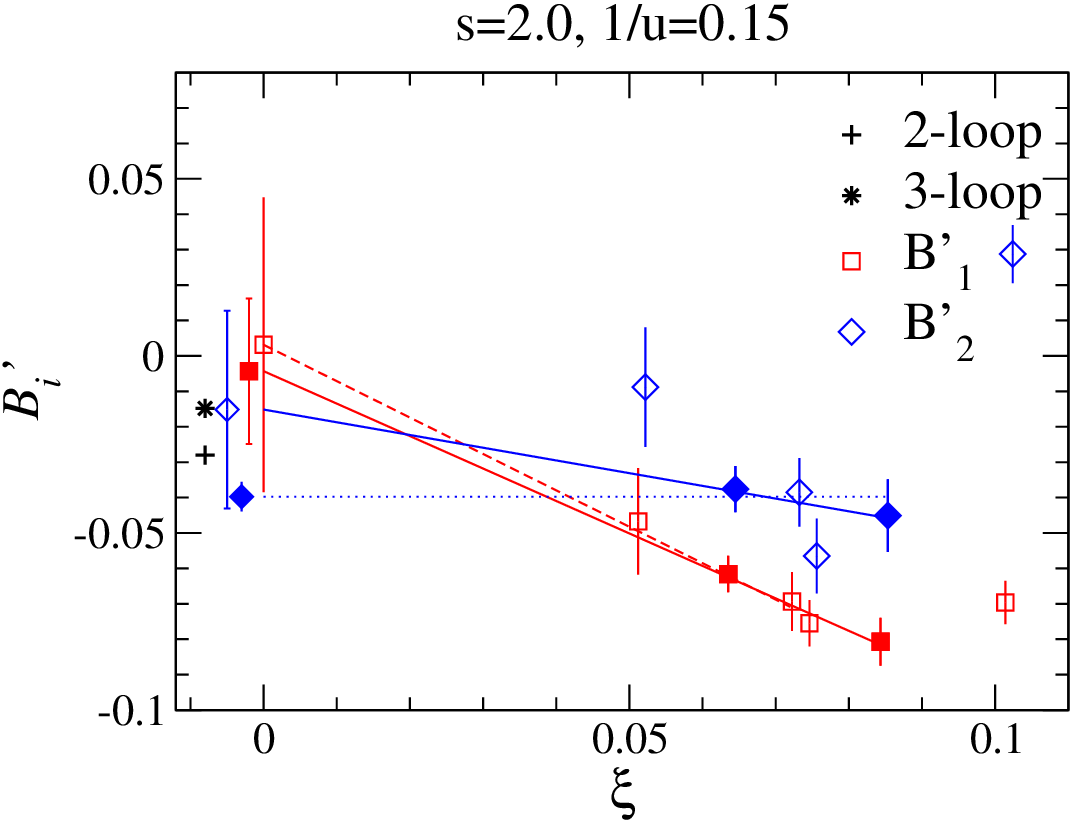}\\
\end{tabular}
\caption{
 Same as Fig.~\ref{fig:contlimit-g2run-each-weak}, but the result at
 the stronger coupling is plotted.
 Solid lines denote the linear extrapolation using the data with $s'=2$
 (filled symbols).
 Dashed line shows the linear extrapolation using the data with the
 three smallest $\xi$.
 }
\label{fig:contlimit-g2run-each-strong}
\end{figure}

Next let us move on to the result at a stronger coupling shown in
Fig.~\ref{fig:contlimit-g2run-each-strong}.
As seen from the figure, first the data of $B'_1$ except for the one
with ($s', l$)=(4/3, 6) (right-most point) remains to align within the
statistical uncertainty.
Thus, the linear extrapolation of $B_1$ is reliable at $1/u=0.15$.
Secondly, the linear extrapolation of $B_1$ and the constant fit of
$B_2$ lead to different continuum DBF.
It appears that the constant fit of $B_2$ is no longer valid and the
linear fit appears to be more reasonable.
Indeed, the linear fit of $B_2$ (solid line and open diamond at $\xi=0$)
turns out to give the consistent limit as shown in the figure.

From the alignment of $B'_1$, we infer that both $\delta B$ and
non-linear scaling violation remain small.
This is consistent with the fact that the continuum DBF obtained by
linear extrapolation of $B_1$ is consistent with zero and thus
$\delta B$ should be small as well.

The deviation of the coarsest data from the linear behavior indicates
that the linear discretization error no longer dominates others in the
data with ($s', l$)=(4/3, 6).
Since in general non-linear scaling violation can be large for small
$l$, the data with $l=6$ may suffer from this though it is not visible
in the figure.
To evaluate the potential uncertainty due to the $O(l^{-2})$
discretization error, we performed a linear fit without the $l=6$ data.
The fit result is shown as open square at $\xi=0$ and the dashed line
in Fig.~\ref{fig:contlimit-g2run-each-strong}.
The result is consistent with that using the $s'=2$ data only.

From Figs.~\ref{fig:contlimit-g2run-each-weak} and
\ref{fig:contlimit-g2run-each-strong}, it turns out that for
$1/u\ \siml\ 0.3$ the extracted continuum DBF is consistent with zero.
This indicates that in this region the running coupling constant reaches
an infrared fixed point or, at least, the running appreciably slows
down.
In order to further investigate the existence of the infrared fixed
point, we include the data obtained from $l$=18 lattice at $\beta=4.55$
into analysis.
This data is combined with the data with $l=12$ to construct $B'_1$ with
$(s', l)$=(3/2, 12).
At $\beta=4.55$, the inverse SF coupling for $l$=12 turns out to be
$1/u=0.107$.
On $l$=6 lattice, this value of $1/u$ is realized at $\beta \sim 4.4$.
In such a small $\beta$, the SF couplings are not calculated on $l$=12,
16 lattices, and hence the following analysis is carried out without the
data from $l$=6 lattices.

\begin{figure}[tb]
\centering
\begin{tabular}{c}
\includegraphics*[width=0.5 \textwidth,clip=true]
{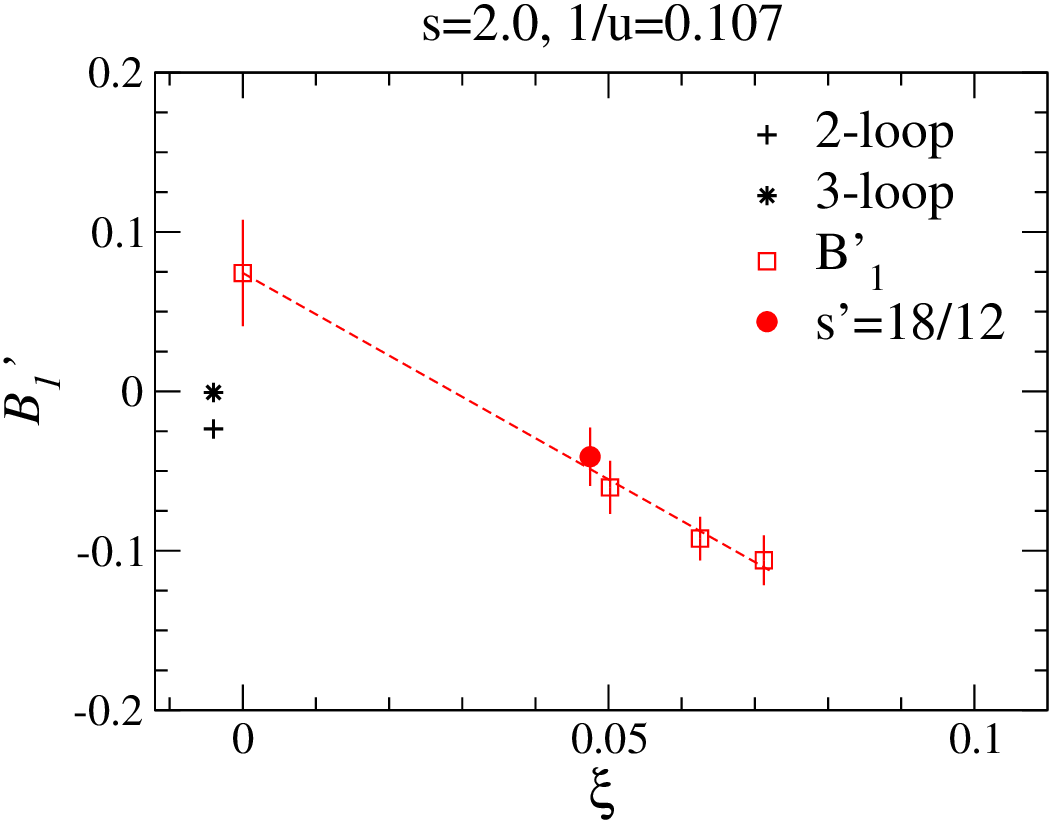}\\
\end{tabular}
\caption{Same as Fig.~\ref{fig:contlimit-g2run-each-strong} but the data
 point obtained with ($s',\ l$)=(3/2, 12) (filled circle) is included in
 the analysis at $1/u$=0.107.
  The dashed line and the open square at $\xi=0$ are the result of the
 linear fit.
}
\label{fig:contlimit-g2run-wt-L18}
\end{figure}

$B'_1$ constructed from the $l$=18 data is shown in
Fig.~\ref{fig:contlimit-g2run-wt-L18} (filled circle).
Since the four data points shown align well, we take the linear
extrapolation using all of them and obtain the positive value in
$\xi=0$.
Interpretation of this result needs care as mentioned
sec.~\ref{sec:remarks}.
The most plausible explanation for this observation is that an IRFP
exists in $u_{\rm IRFP} < 1.0/0.107=9.35$.

\begin{figure}[tb]
\centering
\begin{tabular}{c}
\includegraphics*[width=0.5 \textwidth,clip=true]
{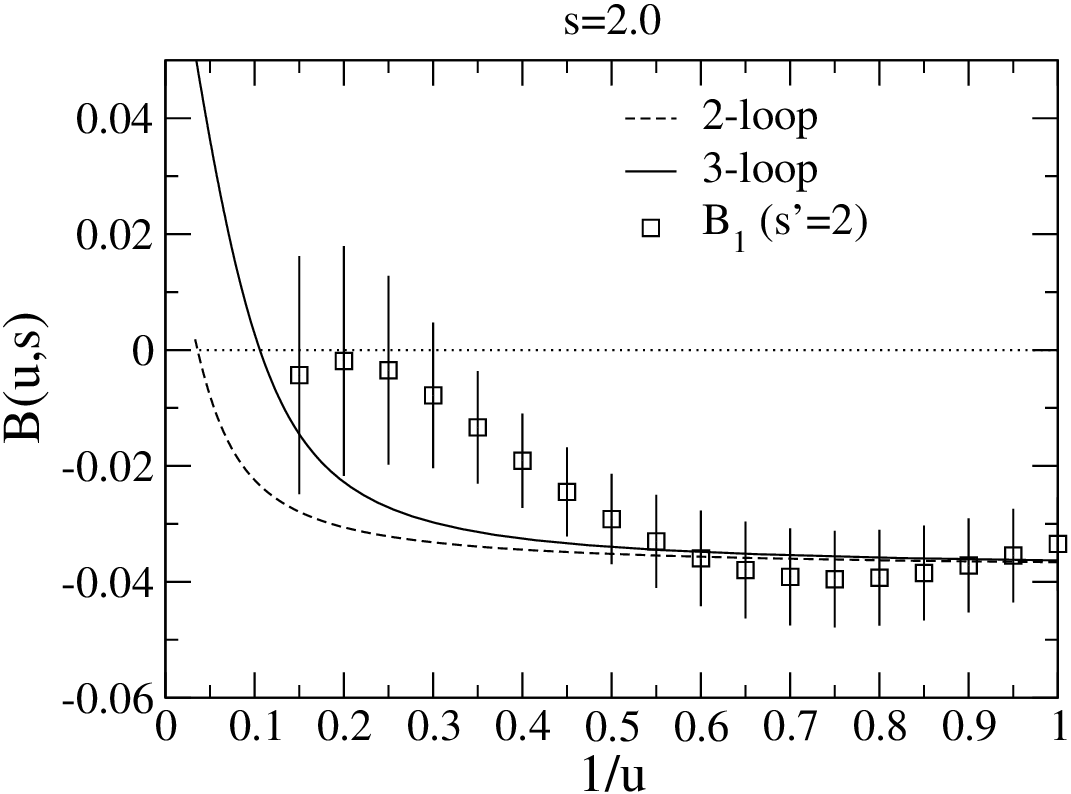}\\
\end{tabular}
\caption{
 $1/u$ dependence of $B(u,s)$ with $s$=2 obtained from the linear
 extrapolation of the data with $s'=2$.
 Two- and three-loop perturbative predictions are shown by dashed and
 solid line, respectively.}
\label{fig:udep-dbf}
\end{figure}

Figure \ref{fig:udep-dbf} shows the $1/u$ dependence of the continuum
DBF, where the results are compared with the perturbative calculations.
It is seen that the running starts to slow down at around
$1/u \sim 0.5$, and eventually the coupling constant reaches a fixed
point in the range of $0.107 < 1/u\ \siml\ 0.3$.
When the DBF is positive, it is non-trivial for the continuum limit to
exist.
Thus we omit the positive DBF data from the figure.

\section{Summary and outlook}
\label{sec:summary}

In this work, the running coupling constant of ten-flavor QCD is
numerically investigated using lattice technique.
The extrapolation of the DBF to the continuum limit is taken linearly
assuming that the $O(a)$ scaling violation dominates the higher order
ones.
The DBF extrapolated approaches zero from below as the SF coupling
constant $u$ increases and when $u \simg 10/3$ the DBF becomes
consistent with zero.
Further investigation at one particular strong coupling $u=9.3$
($1/u=0.107$) is made using the data from the large lattice ($l=18$),
and suggests that the continuum DBF at this coupling is not negative.
This indicates the existence of the infrared fixed point
$10/3\ \siml \ g_{\rm IRFP}^2\ \siml \ 9.3$. 
The linear extrapolation is reasonably justified within the statistical
error, but further rigorous check is clearly preferable.
Combining our result with that of Ref.~\cite{Appelquist:2007hu}, the
critical number of flavors which separates the conformal phase and the
broken phase is $8 < N_f^{\rm crit} < 10$.

In order to confirm the existence of IRFP or even determine the
value of the fixed point more precisely, data from larger lattices with
high statistics are necessary.
It is, however,  difficult to do with machines currently available to
us, and probably more efficient methods or different approaches are
necessary to go further.
As mentioned in sec.~\ref{sec:introduction}, the conformal window can
also be studied by looking at hadrons' spectroscopy or renormalization
group analysis on the lattice.
Currently the conclusions based on various methods are not consistent
among them.
In order to pin down $N_f^{\rm crit}$, these contradictions must be
clarified with further studies.

What is really important in the context of the WTC is the anomalous
dimension of the $\bar \psi \psi$ operator.
The calculation of the anomalous dimension in ten-flavor QCD is
on-going.
The result will be published elsewhere.

Once one has fixed an attractive candidate for WTC, the next important
step would be the calculation of the $S$-parameter.
The calculational method has been established in
Ref.~\cite{Shintani:2008qe}, where the QCD $S$-parameter is calculated
on the lattice for the first time and is correctly reproduced.
Later, the method was applied to three-flavor QCD~\cite{Boyle:2009xi},
sextet QCD~\cite{DeGrand:2010tm} and
six-flavor QCD~\cite{Appelquist:2010xv}.
In Ref.~\cite{Appelquist:2010xv}, the evidence of the reduction of
$S$-parameter is reported.
Another important quantity which should be calculated is obviously the
mass spectrum of the candidate theory, including vector and scalar
resonances, the decay constant of the NG boson and the chiral
condensate.
Although the precise determinations of these quantities are challenging,
the direct comparison with the upcoming LHC results is extremely
interesting and hence we believe that such calculations are worth a lot
of efforts.

\section{Acknowledgment}

N.Y. thanks the Aspen Center for Physics where the workshop ``Strong
Coupling Beyond the Standard Model'' were held during May and June
2010 and the Yukawa Institute for Theoretical Physics at Kyoto
University for supporting the YITP workshop ``Summer Institute 2010''
(YITP-W-10-07).
N.Y. would also like to thank Hideo Matsufuru, Yoshio Kikukawa and
Takashi Kaneko for useful discussion.
A part of this work was completed during these workshops.
The main part of the numerical simulations were performed
on Hitachi SR11000 and the IBM System Blue Gene Solution at High Energy
Accelerator Research Organization (KEK) under a support of its Large
Scale Simulation Program (No. 09/10-01, 10-15),
on B-factory computer system at KEK,
on GCOE (Quest for Fundamental Principles in the Universe) cluster
system at Nagoya University, and
on the INSAM (Institute for Numerical Simulations and Applied
Mathematics) GPU cluster at Hiroshima University.
This work is supported in part by the Grant-in-Aid for Scientific
Research of the Japanese Ministry of Education, Culture, Sports, Science
and Technology
(Nos.
20105001, 
20105002, 
20105005, 
21684013, 
22740183, 
22011012, 
20540261, 
22224003, 
20740139, 
18104005, 
22244018, 
and
227180    
), and by
US DOE grant \#DE-FG02-92ER40699. 

\end{document}